\documentclass[sn-mathphys,Numbered]{sn-jnl}
\usepackage{graphicx}%
\usepackage{multirow}%
\usepackage{amsmath,amssymb,amsfonts}%
\usepackage{amsthm}%
\usepackage{mathrsfs}%
\usepackage[title]{appendix}%
\usepackage{xcolor}%
\usepackage{textcomp}%
\usepackage{manyfoot}%
\usepackage{booktabs}%
\usepackage{algorithm}%
\usepackage{algpseudocode}%
\usepackage{listings}%
\usepackage[makeroom]{cancel}
\usepackage{tabularx,subcaption,fancyhdr,mathtools}
\usepackage{hyperref}
\DeclareUnicodeCharacter{2212}{-}

\usepackage{setspace}
\usepackage{blindtext}

\theoremstyle{thmstyleone}%

\theoremstyle{thmstyletwo}%

\theoremstyle{thmstylethree}%

\newcommand{\beq}{\begin{equation}}
\newcommand{\eeq}{\end{equation}}
\def\bea{\begin{eqnarray}}
\def\eea{\end{eqnarray}}

\begin{document}

\title[Article Title]{\textbf{Probing intractable beyond-standard-model parameter spaces armed with Machine Learning}}

\author[1]{\fnm{Rajneil} \sur{Baruah}}\email{e22soep0012@bennett.edu.in}

\author[1]{\fnm{Subhadeep} \sur{Mondal}}\email{Subhadeep.Mondal@bennett.edu.in}

\author[2]{\fnm{Sunando Kumar} \sur{Patra}}\email{sunando.patra@gmail.com }

\author[3]{\fnm{Satyajit} \sur{Roy}}\email{roy.satya05@gmail.com }

\affil[3]{\orgdiv{Department of Physics}, \orgname{Bangabasi College}, \orgaddress{ \city{Kolkata}, \postcode{700009},  \state{West Bengal}, \country{India}}}

\affil[2]{\orgdiv{Department of Physics}, \orgname{Bangabasi Evening College}, \orgaddress{ \city{Kolkata}, \postcode{700009},  \state{West Bengal}, \country{India}}}

\affil[1]{\orgdiv{Department of Physics, SEAS}, \orgname{Bennett University}, \orgaddress{\street{Greater Noida}, \city{Uttar Pradesh}, \postcode{201310}, \country{India}}}


\abstract{This article attempts to summarize the effort by the particle physics community in addressing the tedious work of determining the parameter spaces of beyond-the-standard-model (BSM) scenarios, allowed by data. These spaces, typically associated with a large number of dimensions, especially in the presence of nuisance parameters, suffer from the curse of dimensionality and thus render naive sampling of any kind---even the computationally inexpensive ones---ineffective. Over the years, various new sampling (from variations of Markov Chain Monte Carlo (MCMC) to dynamic nested sampling) and machine learning (ML) algorithms have been adopted by the community to alleviate this issue. If not all, we discuss potentially the most important among them and the significance of their results, in detail.}

\maketitle

\section{Introduction}
\label{sec:intro}
One of the primary objectives of the particle physics community till now has been to identify a new physics (NP) model that can address the shortcomings of the standard model (SM). Albeit some small excesses reported by the experimental collaborations, we do not have any firm indication of a particular BSM theory as of now. However, accelerator experiments, such as the Large Hadron Collider (LHC) \cite{Evans:2008zzb}, electron-positron collider (LEP) \cite{Brandt:2000xk} and Tevatron \cite{Wilson:1977nk}, along with other experiments studying neutrino oscillation \cite{Super-Kamiokande:2019gzr,T2K:2018rhz}, dark matter \cite{Aprile:2005mz}, flavor violation \cite{Meucci:2022qbh} etc, have accumulated large datasets that can be used to probe the existing BSM theories like never before. The biggest challenge we face in this regard is the intractable parameter spaces. These BSM theories introduce new parameters, often in large numbers. Most of these parameters can take up any numerical value, as long as they obey some theoretical constraints. As a result, one has to deal with a multi-dimensional parameter space that can predict the numerical values of several experimental observables. Finding the regions of the NP parameter space therefore---consistent with all experimental observations simultaneously---becomes a problem with increasing dimensionality. The gradually increasing precision of the experimental observations does not help to improve the situation. 

This data-driven approach of `BSM parameter space exploration' ensures the usage of various statistical methods. The generally accepted method used over the years involves computing some likelihood function, using the data on a set of available observables. More often than not, while combining the data from different experimental collaborations---available as numerical estimates---one assumes a Gaussian likelihood corresponding to each observable. The likelihoods are then combined and depending on the statistical philosophy adopted (frequentist, Bayesian, or hybrid), some optimization or sampling algorithm is run using the combined likelihood. From this point on, the words `parameter space' will mean a normalized or unnormalized posterior distribution of the parameters. Again, conventionally, the unnormalized posteriors are obtained by sampling from them, using some version of Markov Chain Monte Carlo (MCMC) \cite{Speagle:2019ffr,brooks2011eds}. 

While efficient for estimating low-dimensional posteriors, these famously suffer from the \textit{curse of dimensionality} (a term originally coined by Richard Bellman in 1957 in the context of approximation theory \cite{Bellman:DynamicProgramming}).The basic versions are moderately fast but non-parallelizable. Parallelizable versions exist, but there is a trade-off in terms of the increased number of individual likelihood computations; these are problematic for any typical high-energy phenomenological problem, where calculating the likelihood is the main bottleneck in computation time. Moreover, in the presence of potentially disconnected high-likelihood regions (pockets of the parameter space, consistent with experimental observations)---which is often the case---the MCMC chain tends to get stuck in one and often fails to discover the whole parameter space \cite{tripuraneni2017magnetic}. Furthermore, the convergence of the MCMC runs requires quite a large number of points, a priori unknown, which again contributes to the bottleneck mentioned earlier, thus rendering the process computationally expensive.

Nested Sampling (NS) \cite{Skilling:2004,Skilling:2006} is a numerical integration strategy used to calculate the Bayesian evidence, and as a by-product, one can also generate posterior inference. As NS populates the whole posterior together, it is more efficient for multi-nodal posteriors, compared to MCMC. However, as the algorithm moves towards higher likelihood regions, it slows down considerably before converging. Hence, one inevitably requires high computation power to obtain reliable and consistent sampling of the parameter space. {Multinest \cite{Feroz_2009} has been used for performing NS in publicly available global fitting package GAMBIT \cite{GAMBIT:2017yxo,Bloor:2021gtp}.
The use of different sampling algorithms, especially the ones augmented with ML algorithms, has opened up possibilities to improve upon this aspect.


ML algorithms can be broadly divided into two categories; supervised and unsupervised learning. Though unsupervised learning (e.g. clustering) is widely used within particle physics for anomaly detection \cite{Belis:2023mqs}, it has limited applicability in parameter inference. We will only discuss supervised ML in this article, with one example where the method is a modification of an unsupervised technique. ML algorithms, adopted for the sampling of NP parameter space can be categorized into classifiers and predictors. The objective of classifiers is to identify `good' and `bad' points on the parameter space subjected to experimental data. One can use such a classifier to draw a decision boundary as well, that isolates the experimentally allowed region. Predictors, on the other hand, perform regression by matching the predicted outputs to experimental results. In more conventional ML algorithms, training of these classifiers or predictors are done over an infinite pool of data. In algorithms such as active learning the training is performed while finding the favoured parameter space. Active learning is being used widely in the particle physics community in order to study multi-dimensional parameter space \cite{Caron:2019xkx,Goodsell:2022beo}.     

It is an iterative method guided by machine learning to sample specific regions of a multi-dimensional parameter space. This approach is efficient in identifying `good' and `bad' points, i.e., points capable of fitting a certain set of observable within acceptable uncertainties or not. Through multiple iterations, active learning can perform as a better classifier \cite{Caron:2019xkx,Goodsell:2022beo}. The neural network (NN) classifiers themselves give rise to additional uncertainties \cite{DBLP:journals/corr/abs-2107-03342} which need to be taken into consideration. Once the network is properly trained, the user can use the network to predict parameter points that can produce high likelihood or minimize the loss function \cite{Goodsell:2022beo,Hollingsworth:2021sii}. Methods like Hamiltonian Monte Carlo \cite{betancourt2018conceptual} have been used to sample the likelihood \cite{Hollingsworth:2021sii}. Generative models such as normalizing flow \cite{Bishop} have also been used for training purposes \cite{Hollingsworth:2021sii}.

However, it is definitely more informative to obtain the posterior inference on the parameter space rather than just identifying the `good' and `bad' points. To that end, one can pair up NS algorithms with machine learning techniques in a bid to speed up the process. As will be shown later in this draft, it involves training an NN iteratively and letting it predict points for the NS step, with increasing efficiency as the algorithm converges toward the high-likelihood regions. We have implemented this algorithm in our own framework as will be discussed in detail. We also present some results to showcase the efficiency of our framework. 

The article is organized as follows. In sec.~\ref{sec:sampling} we briefly introduce the Bayesian interpretation of probability and the likelihood function. We proceed to introduce the MCMC algorithm and showcase two examples of its wide range of applications. Next, we introduce the idea of nested sampling and discuss why it performs better than MCMC. In sec.~\ref{sec:ML} we summarise some major developments regarding applications of ML. We introduce active learning as a tool to obtain decision boundaries with some examples and then proceed to discuss how higher dimensional parameter spaces can be sampled by careful structuring of the latent space. We then discuss how to use a regressor to reconstruct the likelihood function and how one can make a classifier and a regressor work in tandem to be computationally more effective. In sec.~\ref{sec:MLAssist} we propose an algorithm based on nested sampling assisted by ML in order to generate posterior inference as well. We discuss the algorithm in detail and also showcase some results arguing why this algorithm is computationally more economical in sampling large intractable parameter regions. Finally, in sec.~\ref{sec:summary} we summarise the article.

\section{Conventional statistical sampling algorithms}
\label{sec:sampling}
\subsection{Basics}\label{sec:stat_basics}
The high precision of experimental data and the large number of input parameters in a typical BSM scenario make it a particularly challenging problem to sample the parameter space of the NP model in question. Given the possible wide, theoretically allowed ranges of the observables and the precision of experimental data, a random point scan of the parameter space is not feasible. Only some kind of parameter inference, either maximum likelihood estimation (MLE) using some global optimization technique, or some algorithm that samples the `Posterior' distribution of the parameters, makes sense. We will only talk about the latter in this draft. Let us briefly revisit some definitions in that context.

The Bayesian interpretation of probability is the `degree of belief' or `state of knowledge', which we update using Bayes' theorem:
\begin{equation}
    P(A|B) = \frac{P(B|A) P(A)}{P(B)}
\end{equation}
which changes to the following form in the context of gaining knowledge from data:
\begin{equation}\label{eq:posterior}
    P(\Theta|D) = \frac{P(D|\Theta) P(\Theta)}{P(D)} = \frac{\mathcal{L}(\Theta) \pi(\Theta)}{Z}
\end{equation}
where the prior, $\pi(\Theta) = P(\Theta)$ represents what was known about the parameters of a model before the introduction of the observed data ($D$), the posterior, $P(\Theta|D)$, represents what is known after learning from the data. $\mathcal{L} = P(D|\Theta)$ denotes the \textit{likelihood} of the data in the presence of the model in question. The denominator, $Z = P(D)$, is the \textit{evidence} or \textit{Marginal Likelihood} that appears in Bayesian model comparison \cite{Kass:1995}. It may be written as,
\begin{equation}\label{eq:Z1}
    Z = \int{\mathcal{L}(\Theta) \pi(\Theta)} d\Theta = \int{\mathcal{L}(\Theta) d\mu(\Theta)}\,.
\end{equation}
It normalizes the posterior so that $\int{P(\Theta|D)} d\Theta = 1$. The ratio of the evidences computed for different models is known as a \textit{Bayes factor}, $B_{10} = Z_1 / Z_0$, which updates the relative plausibility of two models in the light of data.

The high energy phenomenology (HEP) community predominantly uses the Gaussian or `Normal' form of the likelihood function, due to the way the experimental results are available to the larger community. With this approximation, the likelihood function is
\begin{equation}
    \mathcal{L} \propto \exp(-\frac{\chi^2}{2})\,.
\end{equation}
where $\chi^2$ is defined as
\begin{equation}
    \chi^2 = \sum_{i=1}^{n_{\rm obs}} \left( O_i^{\rm obs} - O_i^{\rm th}\right)^T \mathcal{C}^{-1} \left( O_i^{\rm obs} - O_i^{\rm th}\right)
\end{equation}
where $O_i^{\rm obs}$ are the set of $n_{\rm obs}$ experimental observables along with associated covariance matrix $\mathcal{C}$ whereas $O_i^{\rm th}$ are the predicted values of the observables obtained from the model.

\subsection{Markov Chain Monte Carlo}\label{sec:mcmc}
MCMC techniques aim to sample the unnormalized posterior by actually sampling from the numerator (product of likelihood and prior) of the right-hand side of eq. (\ref{eq:posterior}). The use of MCMC in HEP has increased by many folds in the last couple of decades, mainly due to progress in computing capabilities. Usages span heavy flavor physics, electroweak precision analyses, BSM physics, and higgs physics. Softwares like HEPfit \cite{DeBlas:2019ehy} and BAT \cite{Beaujean:2011zz}(C++), emcee \cite{Foreman-Mackey_2013} and ptemcee \cite{ptemcee:stv2422} (python), and OptEx \cite{sunando_patra_2019_3404311} (Wolfram Language) are used to perform these analyses. {For some related recent studies, refer to: \cite{Choudhury:2023lbp,Anisha:2020ggj,Jaiswal:2020wer,Bhattacharya:2019der,Biswas:2021pic,Kundu:2021emt,Frank:2021nkq, Castro:2022zpq, Bagnaschi:2022whn, Lu:2022bgw, Karan:2023kyj, Cirigliano:2023nol, deBlas:2021wap, Becirevic:2019tpx, Durieux:2019rbz, Chrzaszcz:2019inj, Bhom:2020lmk, GAMBIT:2023yih, Chang:2023cki, Chang:2022jgo, Balazs:2022tjl, GAMBIT:2021rlp}. To showcase the utilities and shortcomings of a traditional MCMC analysis, we provide two short examples here:

\begin{figure}
    \centering
    \begin{subfigure}{0.43\textwidth}
        \includegraphics[width=\textwidth]{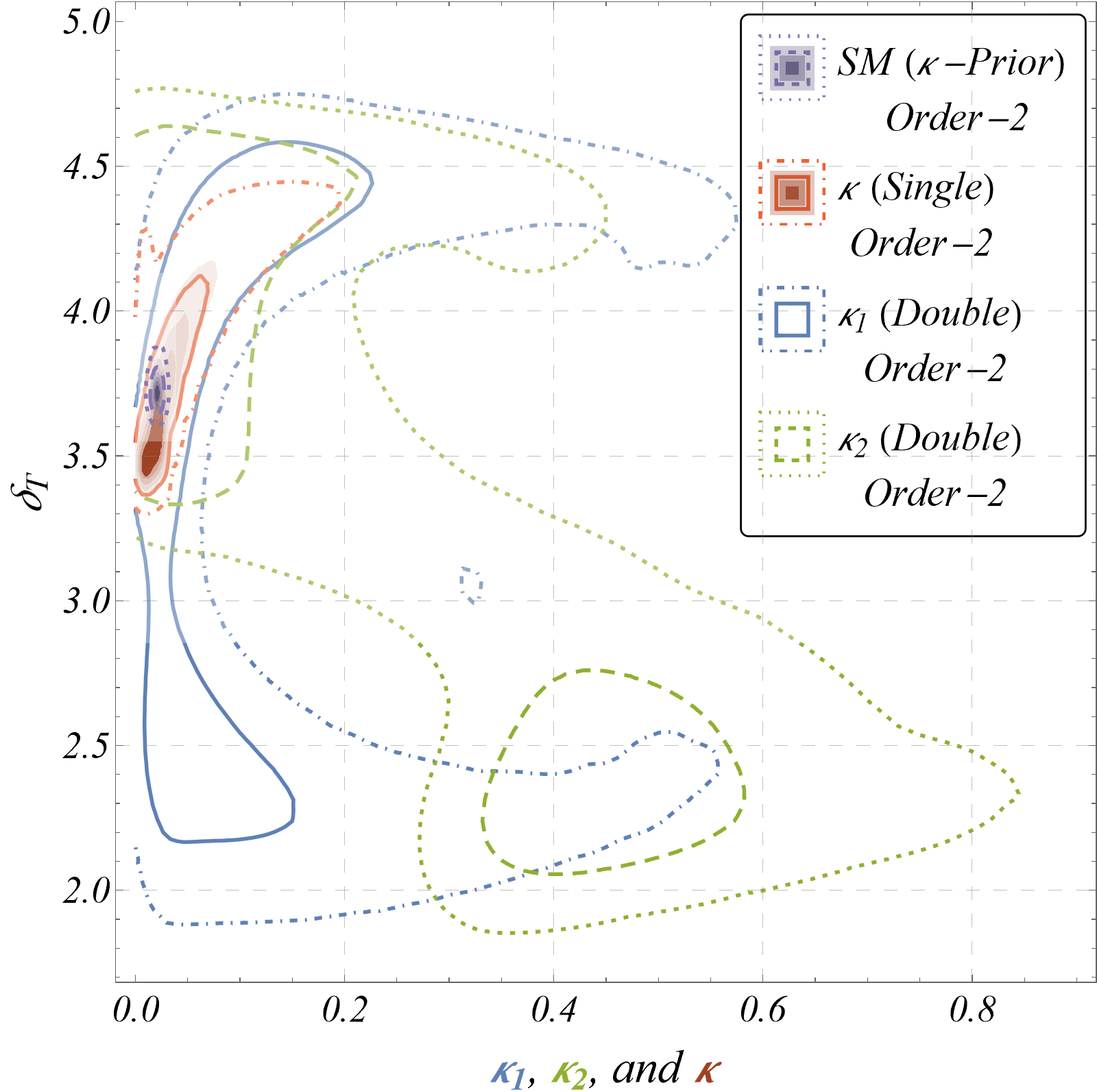}
        \caption{$\kappa_1, \kappa_2$ and $\kappa$ vs. $\delta$}\label{fig:acp1}
    \end{subfigure}\hfill
    \begin{subfigure}{0.45\textwidth}
        \includegraphics[width=\textwidth]{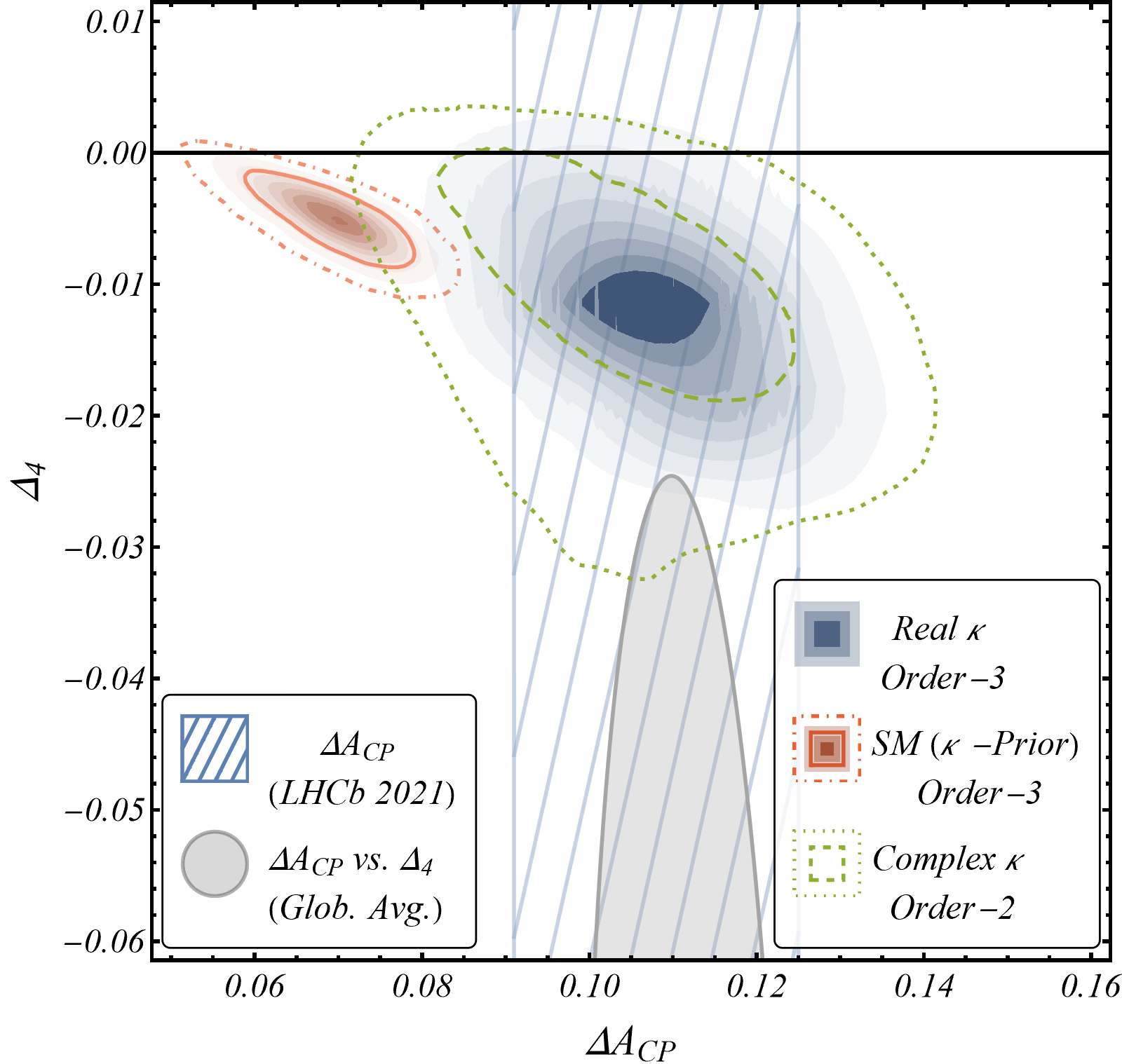}
        \caption{$\Delta A_{CP} - \Delta_4$.}\label{fig:acp2}
    \end{subfigure}
    \caption{Marginal posteriors (2-D) with constant probability contours and the predicted combined distributions of observables $\Delta A_{CP}$ and $\Delta_4$. See sec. \ref{sec:mcmc_ex1}) and ref. \cite{Kundu:2021emt}.}\label{fig:dacp}    
\end{figure} 

\subsubsection{MCMC in analyzing the $\pi-K$ puzzle}\label{sec:mcmc_ex1}
$\Delta A_{CP}$ is the difference in the direct CP-assymmetries between the modes $B^+ \rightarrow \pi^0 K^+$ and $B^0 \rightarrow \pi^- K^+$. A recent claim by LHCb collaboration was in an uncomfortable $8\sigma$ tension with the SM. Investigating this claim including the CP-averaged branching ratios, the authors of ref. \cite{Kundu:2021emt} performed a Bayesian analysis with topological analyses and their phases as the free parameters, to show that the actual deviation of the SM with the experimental result, if any, is still not considerable (within $2\sigma$). 

The modes in question are related by weak isospin and the SM values of the $\Delta A_{CP}$ should be close to zero. This expectation balances on the relative importance of certain flavour-flow topologies and on the assumption that the strong phase difference between the tree and electroweak penguin amplitudes is identically zero. A closer inspection of these topological amplitudes in light of data has been performed using MCMC (OptEx \cite{sunando_patra_2019_3404311}) with Adaptive Metropolis-Hastings algorithm.  Fig. \ref{fig:acp1} shows the 2-D marginal posteriors of several different such parameterizations.

The work also utilizes this sample to compare the predicted $\Delta A_{CP}$ and another such asymmetry ($\Delta_4$), which has to be zero in SM as well. The combined analysis shows that, given the present scenario, $\Delta_4$ is a better probe of BSM effects (fig. \ref{fig:acp2}).

This analysis would serve as an example of some shortcomings of the MCMC method. It searches for some high-density regions in the parameter space, which are allowed by data but non-SM. This is a typical multi-modal problem, where MCMC searches are either hard to tune (e.g., Metropolis, HMC, etc) or are computationally expensive (ensemble methods such as Affine invariant MCMC: emcee). It also compares several different cases, which candidate models to explain the scenario, making it a possible model-selection problem. The authors had to do a hybrid (i.e., a frequentist, point-based, goodness-of-fit) analysis in addition to getting a hold of that information. Calculation of the denominator of eq. \ref{eq:Z2}, the \textit{evidence} would have been a straightforward way to do it.

\begin{figure}
    \centering
    \begin{subfigure}{0.48\textwidth}
        \includegraphics[width=\textwidth]{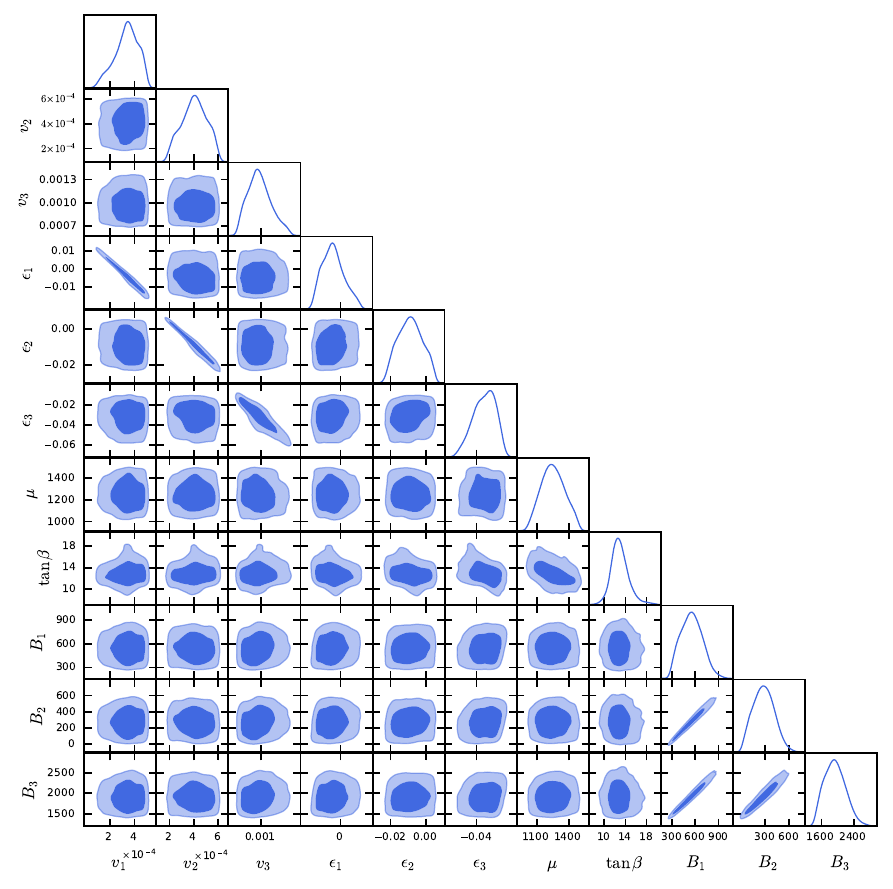}
        \caption{\small 1 and 2-D marginal posteriors in the normal hierarchy.}\label{fig:nh_corner}
    \end{subfigure}\hfill
    \begin{subfigure}{0.52\textwidth}
        \includegraphics[width=\textwidth]{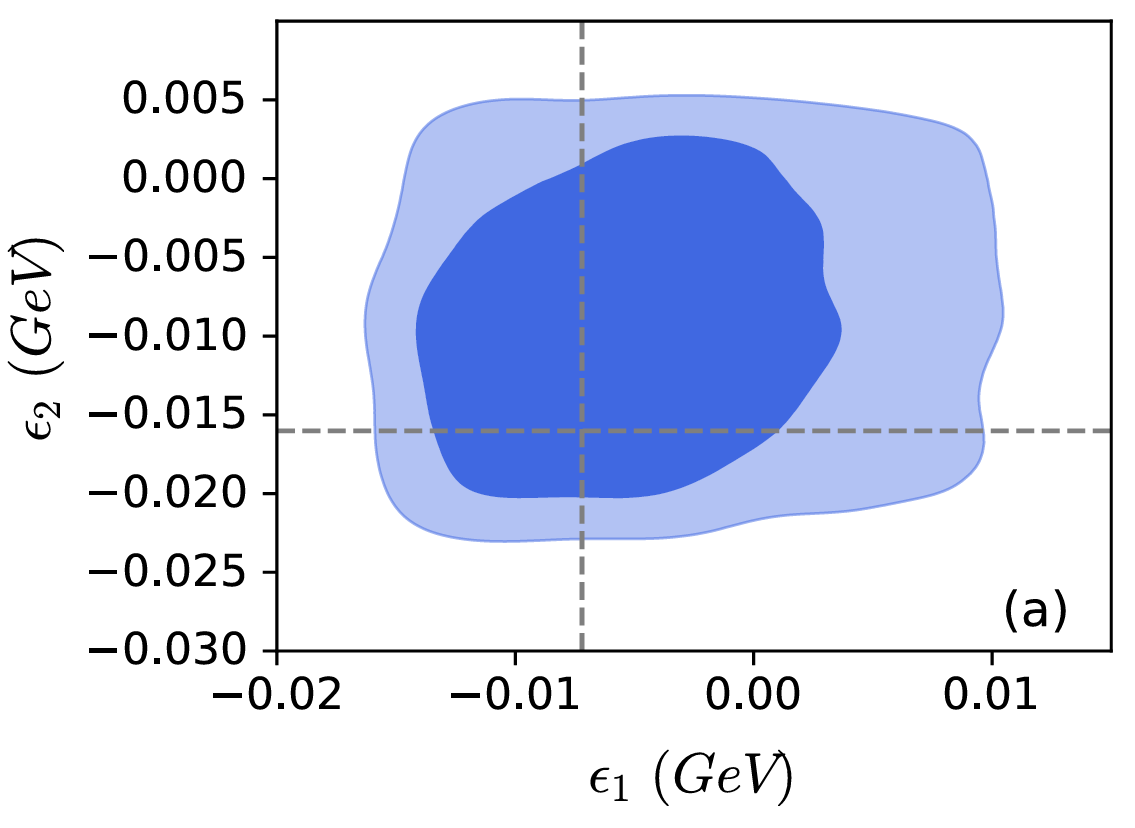}
        \caption{\small The $\epsilon_1 - \epsilon_2$ plane.}\label{fig:eps_mar2}
    \end{subfigure}
    \caption{\small Marginalized posteriors in the normal hierarchy scenario. The dark blue and light blue regions represent contours at 68\% and 95\% C.L. The dashed grey lines indicate the best-fit values for both parameters. See sec. \ref{sec:mcmc_ex2} and ref. \cite{Choudhury:2023lbp}.}\label{fig:normal_hierarchy} 
\end{figure} 

\subsubsection{{MCMC in Supersymmetric parameter spaces}}\label{sec:mcmc_ex2}
MCMC was used in a recent study performed to sample the parameter space of the R-parity violating (RPV) supersymmetric (SUSY) scenario in light of updated light neutrino oscillation data \cite{Choudhury:2023lbp}. Neutrino oscillation experiments provide measurements of two mass-squared differences, three mixing angles, and one phase. Apart from the phase, the other five parameters have been quite precisely measured \cite{deSalas:2020pgw}. The precision is only expected to grow further. Any complete NP model, therefore, must be able to explain the neutrino oscillation data. The RPV SUSY scenarios are quite well-motivated from the aspect of light neutrino mass generation \cite{Grossman:1997is,Rakshit:2004rj,Grossman:2003gq}. The bilinear RPV SUSY scenario introduces mixing between neutrino and neutralino states, which results in the generation of non-zero mass of one light neutrino at the tree level itself. There are further contributions arising at one loop resulting in non-zero masses of the other two light neutrinos. Considering both these contributions together one can easily explain the neutrino oscillation data \cite{Grossman:1997is,Rakshit:2004rj,Grossman:2003gq}. Apart from the five neutrino oscillation observables (ignoring the phase, which is not very precisely measured yet) the other relevant observables in the context of bilinear RPV SUSY are the 125 GeV higgs boson mass, its signal strength measurements \cite{cms_web1} and some flavor observables \cite{Archilli:2017xmu,HFLAV:2019otj,LHCb:2021vsc}. A detailed analysis in the context of this BSM scenario has been performed \cite{Choudhury:2023lbp} with a total of 15 observables and 11 input parameters. 

SPheno \cite{Porod:2003um,Porod:2011nf} and SARAH \cite{Staub:2008uz,Staub:2010jh,Staub:2015kfa} were used for model calculation and implementation in the spectrum generator. The emcee \cite{Foreman-Mackey_2013} framework was used to incorporate the MCMC algorithm. The analysis was performed with both light neutrino mass hierarchical scenarios. The Fig.~\ref{fig:nh_corner} shows the 1-D and 2-D marginal posterior distributions of all the input parameters. For the 2-D distributions, the dark blue and light blue regions represent $1\sigma$ and $2\sigma$ allowed regions respectively. For the results corresponding to inverted hierarchy and more details, refer to \cite{Choudhury:2023lbp}. One can look at the 2-D marginalized posterior distributions in different coupling planes in order to obtain the $1\sigma$ and $2\sigma$ bounds on the respective RPV couplings. As an example, a sample distribution in the plane of two bilinear RPV couplings $\epsilon_1$ and $\epsilon_2$ is shown in Fig.~\ref{fig:eps_mar2} \cite{Choudhury:2023lbp}.

Although effective, this analysis proves to be computationally expensive. It also faces the hurdle typical to parameter search problems: the existence of undiscovered modes. One can never be completely sure that the parameter space obtained through this analysis is the only possible solution. Similar to the first example, MCMC is not good at identifying multi-modal solutions. Though this used an ensemble technique, the starting point of the random walkers holds undue importance and thus is another cog in need of tuning. 

As we showcased in these two examples, the need of the hour is a faster algorithm that can identify all possible solutions (pockets of parameter space in different regions) simultaneously. 

\subsection{Nested Sampling}\label{sec:ns_description}
Nested Sampling (NS) \cite{Skilling:2004,Skilling:2006} is a sampling algorithm, introduced by John Skilling in 2004, in the context of Bayesian inference, and computation\footnote{For a nice introduction to Nested Sampling, targeted to Physical Scientists, see \cite{Ashton:2022}}. As mentioned earlier, traditional MCMC methods mainly sample the un-normalized posterior of the model parameters, because the marginal likelihood or evidence of the model in question---an intractable integration in most cases, especially challenging in higher dimensions---cannot be calculated. The NS was initially developed to solve these integrals. The field of Cosmology immediately embraced it as it partially overcomes some major difficulties in MCMC, mentioned in the introduction of this writing. Before explaining the advantages in detail, let us get some idea about the method. 

NS is essentially an algorithm for integration but has seen most of its usage in  Bayesian inference, so let us borrow our language from that domain. 

\begin{figure}[h!]
    \centering
    \includegraphics[width=\textwidth]{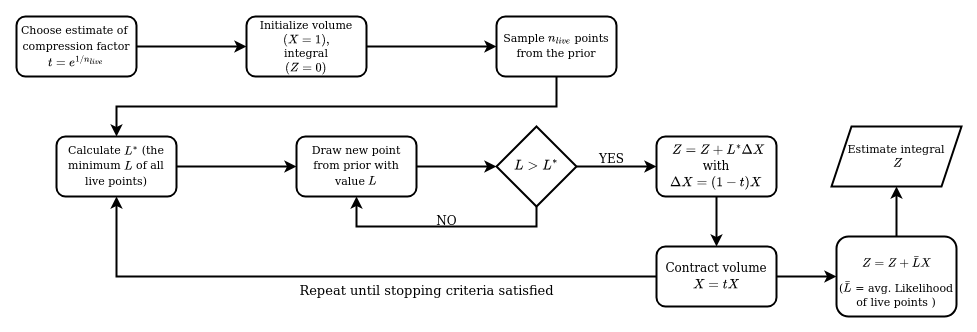}
    \caption{\textit{Schematic of the NS algorithm} for the general multi-dimensional integral.}
    \label{algo:NS}
\end{figure}


NS essentially changes the evaluation of this multi-dimensional integral to a single-variable one, by peeling off equal-probability-contours from the parameter space: 
\begin{equation}\label{eq:Z2}
    Z = \int^1_0{\mathcal{L}(X)} dX \,.
\end{equation}
where $X$ is the volume enclosed by contour $L^*$,
\begin{equation}\label{eq:Z3}
    x(L^*) = \int_{L>L^*}{\pi(\Theta)} d\Theta \,.
\end{equation}
The algorithm depicted in Fig.~\ref{algo:NS} details the general working procedure of NS in pseudocode. For more details, see ref. \cite{Ashton:2022}. Here are the main advantages of using NS:
\begin{itemize}
    \item Developed mainly to calculate $Z$, NS simultaneously returns results for model comparison and parameter inference, hitherto not possible using classical MCMC techniques.
    \item It is successful in multi-modal problems. It samples from the whole prior space simultaneously and can always sample all the modes together except for some pathological cases.
    \item It is naturally self-tuning with far less monitoring, permitting it to be applied immediately to new problems.
    \item As different runs of the same inference problems can be combined to get results with higher precision, NS is essentially parallelizable. This solves many problems faced by the standard MCMC techniques.
\end{itemize}

There have been many software and packages, most in recent years, developed for doing Nested Sampling and variations of it. Multinest \cite{Feroz:2008,Feroz:2009}, CosmoNest \cite{Mukherjee_TAJ:2006,Mukherjee_PRD:2006}, and Dynesty \cite{Dynesty:2020} are some important examples. For a considerably large list of such software, check out Table 2 of ref. \cite{Ashton:2022}.
 
\section{ML as a sampler of parameter spaces}
\label{sec:ML}
Computing a likelihood function from selected observables and model prediction is the conventional way to sample a new physics parameter space. While in theory, these likelihood functions should be continuous, they often act like the step or delta functions when multiple observables are integrated. In this section, we discuss a few methods that guide users in identifying points within the allowed region of parameter space.

\subsection{Active Learning to construct decision boundaries}\label{sec:activelearn}
An algorithm can be developed using Active Learning (AL) \cite{settles:tr09} of the parameter space, where a neural network discriminator is used to identify points close to the \textit{decision boundary}. AL, an iterative method, improves the resolution of the true boundary with each iteration while reducing the runtime of expensive computations (see fig. \ref{fig:al_decision}). Active Learning is a subset of ML algorithms in which the ML algorithm is not a passive participant in a labeled dataset. In such an instance, important points are selected to train the ML algorithm from a pool of unlabelled data. 

The approach used in Active Learning is called the \textit{pool-based sampling.} A flowchart for the same is shown in Fig.~\ref{wrap-fig:1}. 
\begin{itemize}
    \item An initial dataset is curated from the parameter space and passed through the program that `calculates' the label (we will call it the \textit{oracle} from now on) to label it.
    \item An ML \textit{estimator/ algorithm} is trained on this labelled dataset. 
    \item Some more points from the parameter space are sampled and the ML estimator predicts the output. The points with the maximum uncertainty (in the case of a classification problem like this), i.e., the points that are predicted to be at the boundary are then chosen to pass through the oracle and the steps are repeated. 
    \item The hyper-parameters for such a procedure include the size of the initial dataset, the size of the pool(ed) dataset, and the number of points chosen from the pool data to be passed through the oracle.
\end{itemize}

\begin{figure}
    \begin{minipage}{0.48\textwidth}
    \centering
    \includegraphics[width=\textwidth]{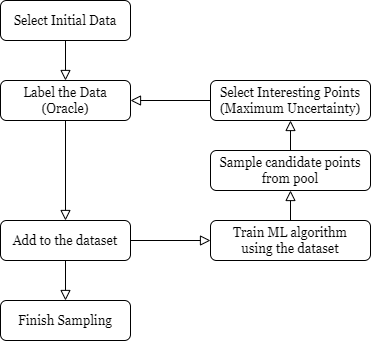}
    \caption{\small Schematic representation of an AL algo. Figure taken from ref. \cite{Caron:2019xkx}.}\label{wrap-fig:1}
   \end{minipage}\hfill
   \begin{minipage}{0.48\textwidth}
     \centering
     \includegraphics[width=\textwidth]{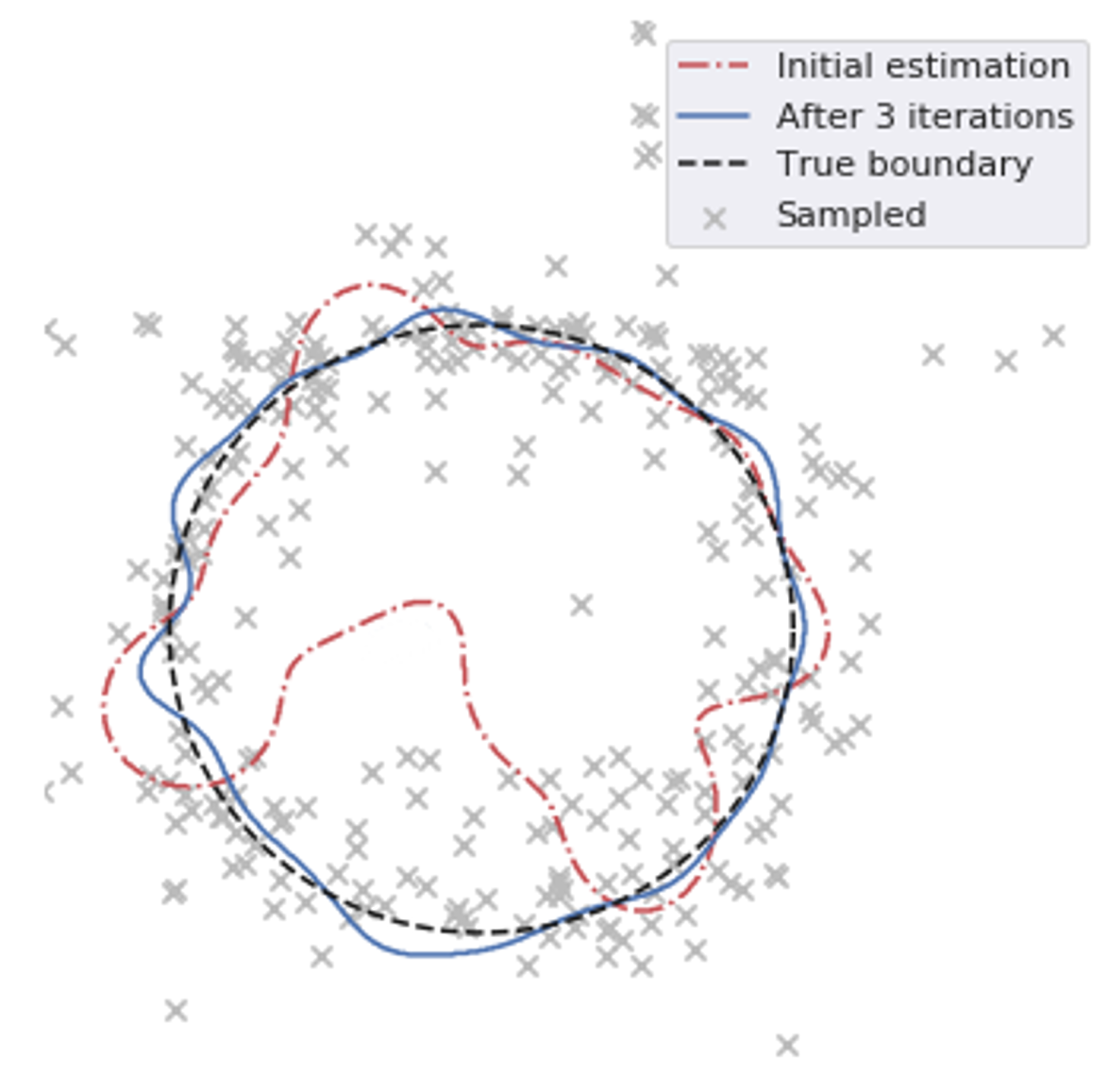}
     \caption{\small Initial estimation (dashed-dotted red line) of the decision boundary (dashed black line) denotes the location with the most uncertain classification. An iterative sampling of new points (crosses) makes new decision boundaries increasingly more accurate: shown for 3 iterations (solid blue line). Figure taken from ref. \cite{Caron:2019xkx}.}\label{fig:al_decision}
   \end{minipage}
\end{figure}

To measure the uncertainty, one can use an algorithm like the Gaussian Process \cite{GaussianProcess1,GaussianProcess2}. However, its computational expensiveness makes it limited to low-dimensional parameter spaces \cite{DBLP:journals/corr/abs-1809-11165}. Two important ways of measuring this uncertainty are Query-by-Committee(QBC) \cite{ArticleByCom} and Query-by-Dropout-Committee(QBDC) \cite{DBLP:journals/corr/DucoffeP15}. IN QBC, multiple ML estimators are trained using the same dataset and their disagreement on the predicted value is used as a measure of uncertainty. In such a case the disagreement between the different estimators can be quantified by the standard deviation. This approach is not bound to a specific type of ML estimator. However, using this approach is not advisable when the committee consists of multiple Neural Networks(NN) as NNs are computationally more expensive. In such cases, QBDC is used where one builds a committee using Monte Carlo dropout \cite{DBLP:journals/corr/DucoffeP15}. Conventionally, dropout layers are used to regularize (prevent overtraining of) the network, and these layers are disabled during test time. However, in Monte Carlo dropout, these layers are enabled even during the test time, such that the output of the NN varies at each evaluation. 

This technique has been used with many different types of ML estimators with different approaches to measure the uncertainty, some of the interesting results \cite{Caron:2019xkx,Goodsell:2022beo} are discussed here.

\subsubsection{pMSSM:}
\paragraph{Using Random Forest with an Infinite Pool}
To have an infinite pool size, the AL algorithm (here: Random Forest) must be given access to a method that can generate new data points. To do that, a trained NN is used as the oracle. The parameter space considered here is the MSSM with 19 independent variables. The observables used here include electroweak precision observables, $\left( g-2\right)_{\mu}$, flavor observables, dark matter relic density, Higgs mass, etc. \cite{Caron:2019xkx}. The accuracy development of AL over Random Sampling can be seen in Fig.~\ref{fig:pmssm_randomforest} \cite{Caron:2019xkx}. The performance gain is defined as the ratio of the number of points needed to reach the maximum performance of Random Sampling over AL. The gain here is between 6 to 7. The bands in the figure correspond to the ranges of accuracy over 7 independent runs.

\begin{figure}[h!]
    \centering
    \begin{subfigure}{0.48\textwidth}
        \includegraphics[width=\textwidth]{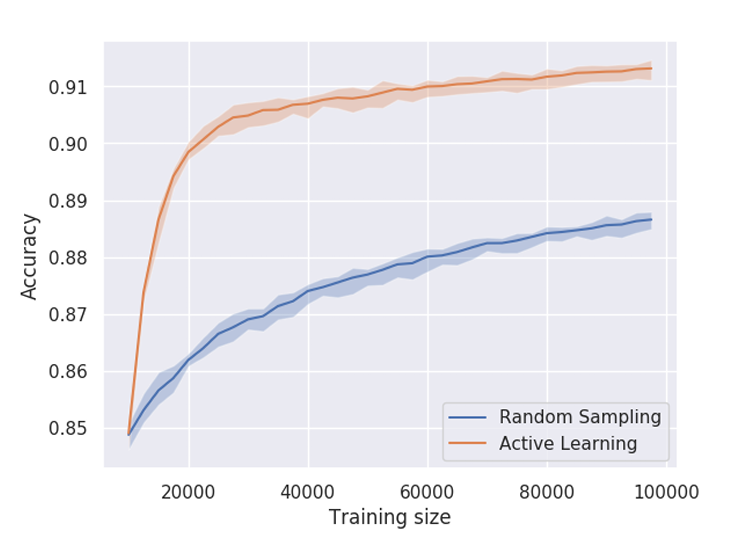}
        \caption{\small Random Forest}\label{fig:pmssm_randomforest}
    \end{subfigure}\hfill
    \begin{subfigure}{0.48\textwidth}
        \includegraphics[width=\textwidth]{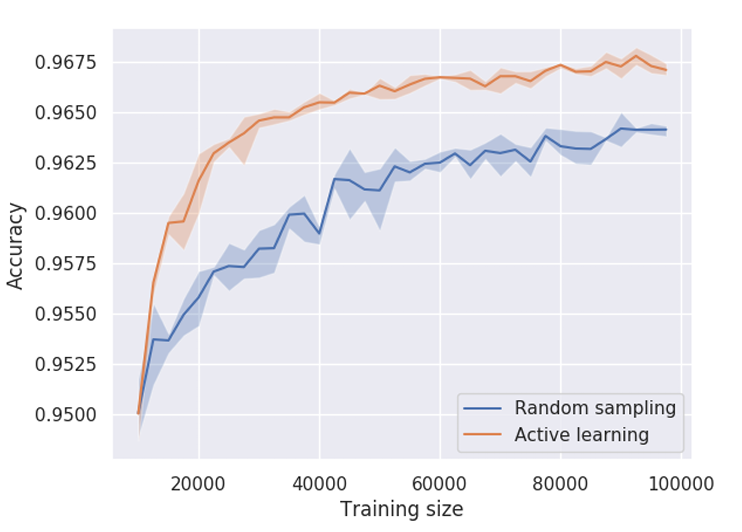}
        \caption{\small Neural Network}\label{fig:pmssm_NN}
    \end{subfigure}
    \caption{\small Accuracy Plot for finding decision boundary of pMSSM using AL with different ML structures with infinite pool taken from Fig.4 and 5 of \cite{Caron:2019xkx}.}\label{fig:pmssm} 
\end{figure}

\paragraph{Using Neural Network with an Infinite Pool}
In this approach, instead of Random Forest, a Neural Network with an infinite-sized pool has been used. The model parameters and the observables remain the same as in the previous case.  The performance gain in this case is around 3 to 4. This method is around K times faster than the previous case with K estimators. We can see in Fig.~\ref{fig:pmssm_NN} \cite{Caron:2019xkx} that compared to the previous case, the accuracy is significantly higher. The reason behind this can either be because NN inherently captures exclusion function better than random forests, or that the architecture of this NN is very similar to the architecture of the oracle. Similar to the previous case, the bands in the figure correspond to the ranges of accuracy over 7 independent runs.

\paragraph{Subspaces of pMSSM}
To investigate the efficiency of this framework while studying lower dimensional parameter spaces, the 19-dimension pMSSM parameter space is somewhat simplified. For comparison, the dimensionality is reduced to 14, 10, 8, 4, and 1 by fixing some soft mass parameters, trilinear couplings, and (or) $\tan\beta$. So that none of the hyperparameters is most effective for active learning, the size of the initial dataset, iteration rate, sample size, and maximum size are all scaled down by the following factor:
\begin{equation}
    \text{scaling} = \left( \frac{\text{number of free parameters}}{19}\right)^2 \nonumber
\end{equation}
The performance of AL with 19, 8, and 1 free parameter(s) is shown in Fig.~\ref{fig:scaling} \cite{Caron:2019xkx}.

\begin{figure}[!ht]
    \centering
    \includegraphics[width=\linewidth]{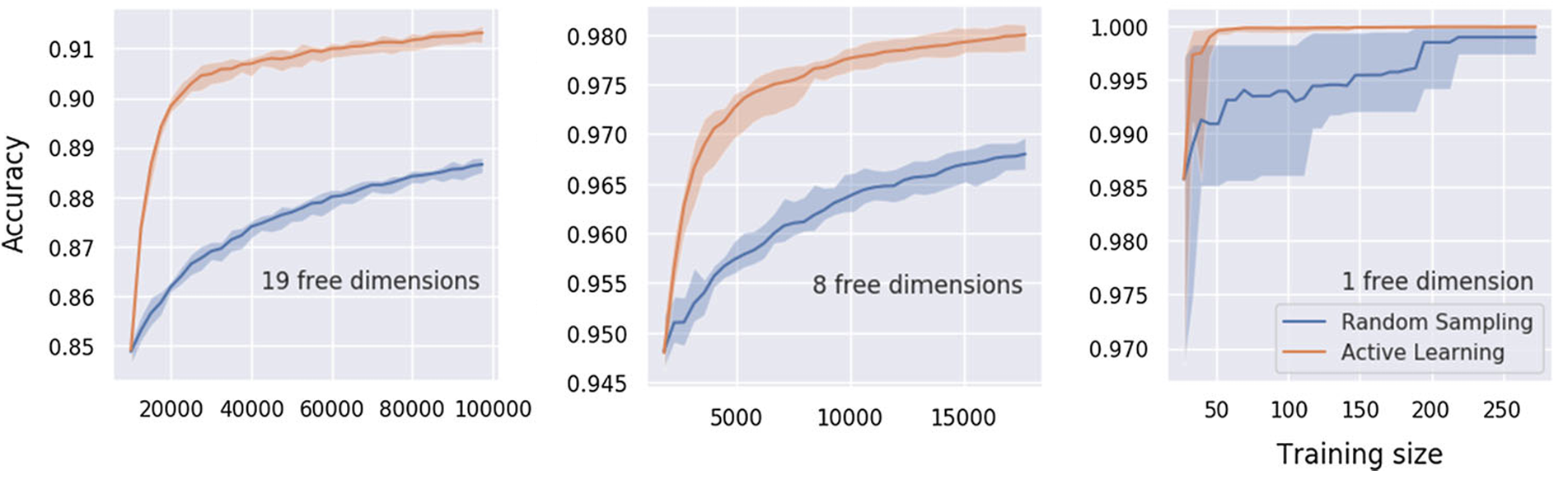}
    \caption{Accuracy plots for AL with Random Forest for different dimensionalities taken from Fig.6 of \cite{Caron:2019xkx}.}
    \label{fig:scaling}
\end{figure}

The relative gain of AL over random sampling is approximately equal in each case. From the plots in the figure, it can be seen that as the dimensionality decreases, the accuracy increases for both AL and random sampling. However, in each case, the accuracy of AL is significantly better than random sampling, that too with a smaller training size. As in the previous cases,  the bands show the range of accuracy when the experiment was repeated 7 times. This shows that the superior performance of AL over Random Scanning remains unaffected in lower-dimensional spaces. However, for random sampling, the accuracy improves significantly as expected as the dimensionality of parameter space decreases.

\subsubsection{cMSSM:}
Another similar attempt to apply AL has been made in ref. \cite{Goodsell:2022beo} to sample points near the decision boundary and visualize them. For the ML estimator, a NN has been used with 2-5 hidden layers, each with an order of 100 neurons that are connected to the ReLU activation function and are finally passed on to a single output neuron with sigmoid activation. To sample interesting points a customized scoring system has been implemented. Fig.~\ref{Fig:Data2} shows the results of a toy modelin 2-D, depicting that AL can identify multiple disconnected regions effectively.

\begin{figure}[h!]
    \centering
    \begin{subfigure}{0.48\textwidth}
        \includegraphics[width=\textwidth]{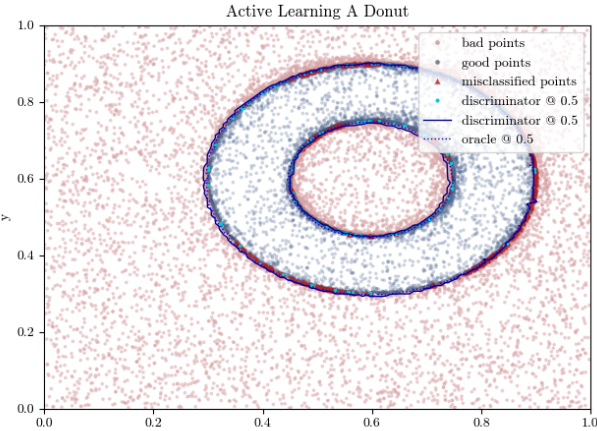}
        \caption{\small Donut shaped parameter space (from fig. 1 in ref.~\cite{Goodsell:2022beo})}\label{Fig:Data2}
    \end{subfigure}\hfill
    \begin{subfigure}{0.52\textwidth}
        \includegraphics[width=\textwidth]{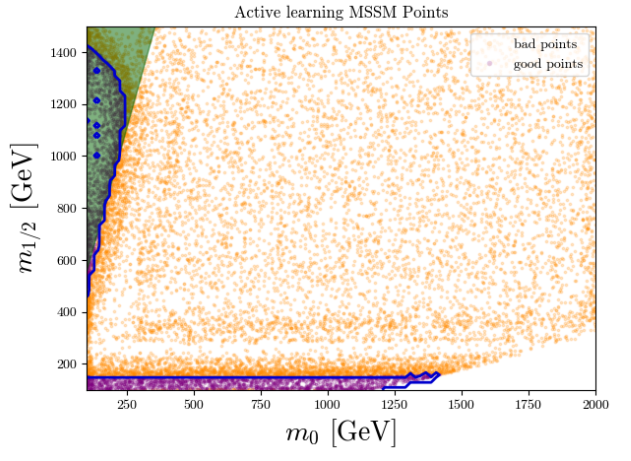}
        \caption{\small AL using NN in cMSSM (from fig. 3 in ref.~\cite{Goodsell:2022beo})}\label{mssm}
    \end{subfigure}
    \caption{\small Performance of AL in sampling complicated regions of parameter space, shown in both a toy and a real BSM example (with the modified scoring formula).}\label{fig:cmssm} 
\end{figure}



Subsequently, the algorithm is tested on constrained MSSM (cMSSM) where the masses and gauge couplings unify at the GUT scale, reducing the number of independent parameters to five: $m_0$, $m_{\frac{1}{2}}$, $A_0$, $tan(\beta)$ and $sign(\mu)$. To resemble the toy problem, three of the five parameters are fixed with the choices $tan(\beta) = 10$, $A_0 = 0$ and $sign(\mu) = 1$. The other two parameters are scanned over a substantial range; $m_0$, $m_{1/2}$ $\in$ $[100, 2000]$ GeV. While the AL scan favors the regions near the decision boundaries, it also explores the rest of the parameter space in sufficient detail (fig.~\ref{mssm}). 

The effectiveness of AL scans is evident in comparison with random scans and traditional MCMC techniques. When pitted against other ML algorithms, these can have smaller structures as new training points are fed into them in a piecemeal fashion, without any loss of accuracy. They also perform better in regions of higher entropy, i.e., more training happens in more uncertain regions, automatically. Sampling multiple disconnected regions also happens organically in these algorithms.

\subsection{Sampling of high dimensional parameter spaces}
\label{sec:higherdim}
\begin{figure}[h!]
    \begin{subfigure}{\textwidth}
    \centering
        \includegraphics[width=0.7\textwidth]{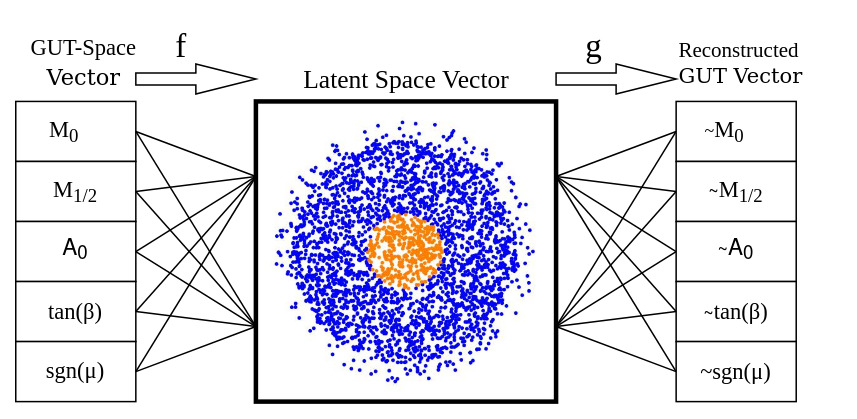}
        \caption{\small The framework for cMSSM. Orange(blue) points are valid(invalid) points.}
    \label{SAFESPAM_schematic}
    \end{subfigure}\\
    \begin{subfigure}{\textwidth}
    \centering
        \includegraphics[width=0.6\textwidth]{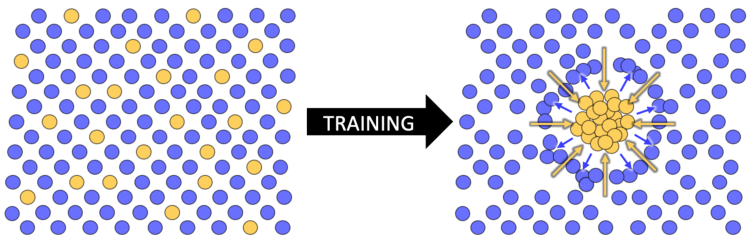}
        \caption{\small Desirable points (orange), all cluster in one region.}
    \label{schematic2}
    \end{subfigure}
    \caption{\small Schematic diagrams of the method discussed in sec. \ref{sec:SWAE}. The figures are taken from ref.~\cite{Baretz:2023mra}}\label{fig:latent} 
\end{figure}

\subsubsection{SWAE to generate structured \textit{Latent Space}}\label{sec:SWAE}

\paragraph{Schematic of the method}
Sampling points from a high-dimensional parameter space can also be achieved by some specific types of dimension-reduction, where ML is used to map the higher dimensional parameter space to a lower dimensional latent parameter space. Some examples of such analyses can be found in refs \cite{Mutter:2018sra, He:2022fxp,Baretz:2023mra}. One can then visualize and explore the lower dimensional latent space to look for clusters of points, in agreement with some experimental dataset. 

However, most of these methods are unsupervised and the resulting latent space often does not have any proper structure. As a result, sampling the `data-allowed' parameter space becomes difficult \cite{DBLP:journals/corr/abs-1804-01947}. One can overcome this issue by generating the lower dimensional parameter space with the help of generative methods through supervised learning. The maps between the parameter space and the latent space should be a) \textit{Generative}: capable of generating new points with higher efficiency; b) \textit{Contrastive}: valid points must be distinctly separated from the invalid ones; c) \textit{Low-Dimensional}: so that sampling and visualization is easier.

To this end, a new class of generative method has been developed \cite{Baretz:2023mra} using Slices Wasserstein Autoencoder (SWAE) structures \cite{DBLP:journals/corr/abs-1804-01947}. The schematic shown in fig.~\ref{SAFESPAM_schematic} takes the cMSSM scenario as an example. The autoencoder maps the GUT scale parameters to a low-dimensional latent space with two parameters that use these structures to treat the problem as a supervised learning task (using the experimental constraints). The SWAE structures are used to create bidirectional maps between the GUT-scale parameter space and the low-dimensional latent space. The structure of this latent space is specialized, in which the \textit{valid points} are forced to cluster near its origin so that new points in the GUT-scale parameter region can be efficiently sampled. The mapping ${f: T \longrightarrow L}$ from the GUT-scale theory space to the latent space is learned by the \textit{encoder}, while the return mapping ${g: L \longrightarrow T}$ is learned by the \textit{decoder}.

\begin{figure}[!ht]
    \centering
    \includegraphics[width=0.8\textwidth]{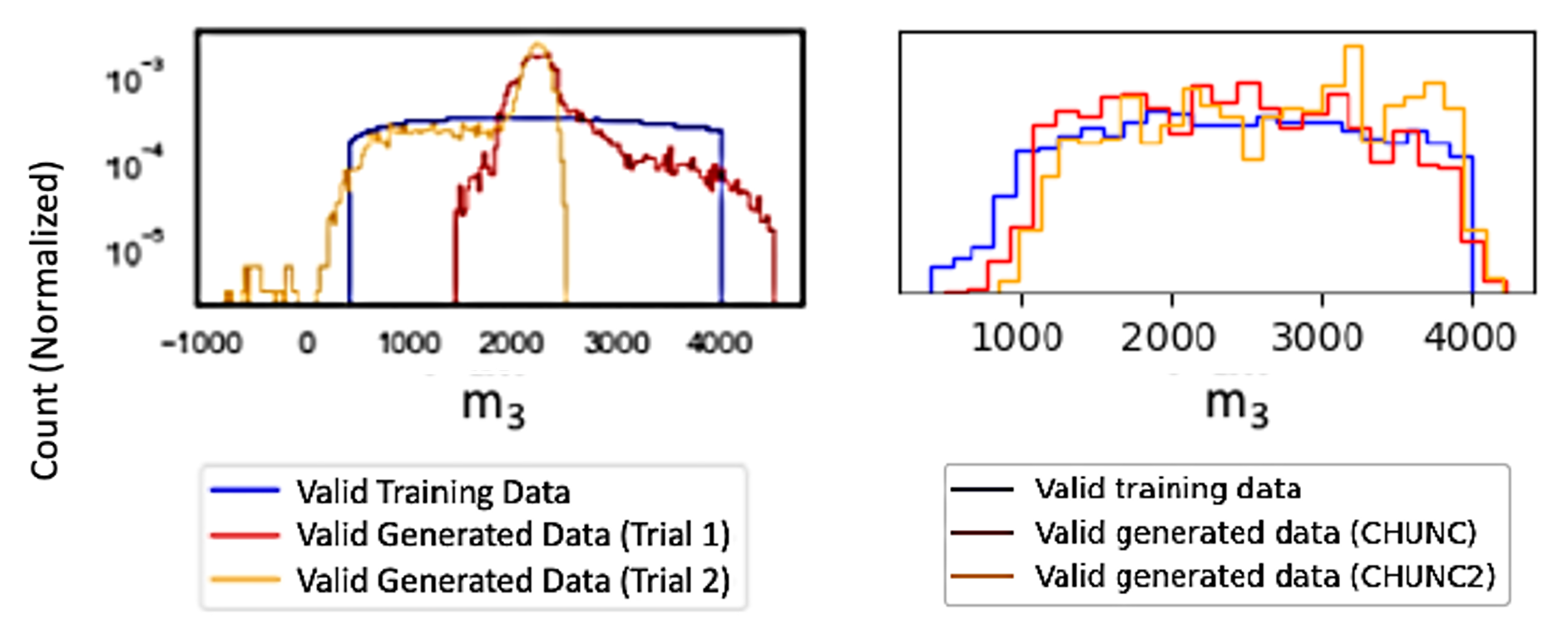}
    \caption{\small Left: Performance of SAFESPAM on the pMSSM parameter $m_3$ using the Higgs Mass as the constraint. The Latent Space is 2 Dimensional. Right: the same for  CHUNC and CHUNC2 using the Higgs Mass, Dark Matter Relic Density and whether or not LSP is the neutralino, as the constraint. Figures are taken from ref.~\cite{Baretz:2023mra}.}
    \label{SAFESPAM_results}
\end{figure}

\paragraph{SAFESPAM}
The primary search algorithm is called the SWAE Algorithm For Experimentally Sound Point-Production and Mapping (SAFESPAM). 
The loss function used in this method, responsible for the structure in the latent space, is a special one:
\begin{equation}
    \mathcal{L} = \alpha \mathcal{L}_{L2} + \beta \mathcal{L}_{Wasserstein} + \gamma \mathcal{L}_{Cluster}
\end{equation}
where $\alpha, \beta, \gamma$ are hyperparameters and,
\begin{eqnarray}
    \mathcal{L}_{L2} &=& \frac{1}{N} \sum_{i=1}^N \left( t_i - g \left( f \left( t_i\right)\right) \right)^2 \\
    \mathcal{L}_{Wassterstein} &=& \frac{1}{MN} \sum_{i=1}^N \sum_{j=1}^M tc \left( \theta_i . \tilde{s}_{k[j]}, \theta_i . f( t_{\ell [j]}) \right)\\
    \mathcal{L}_{Cluster} &=&
    \frac{1}{N} \sum_{i=1}^N \left(a \delta_{valid} ||f(t_i)|| + \frac{b(1-\delta_{valid})}{||f(t_i)||} \right)
\end{eqnarray}
where, $M$ and $N$ are the number of points per batch, $\theta_i$ is the randomly sampled one-dimensional planes along which marginal distributions are defined, $\tilde{s}_k$ and $f( t_{\ell [j]})$ respectively are random samples from the target distribution and the latent space input data. $tc(.,.)$ is a representative of the \textit{transport cost} between both the distributions. $a$ and $b$ are hyperparameters and $\delta_{valid}$ is 1 for valid points and 0 for invalid points. $\mathcal{L}_{L2}$ makes sure that the decoder \textit{accurately} reconstructs the original GUT-scale inputs when mapping.
$\mathcal{L}_{Wassterstein}$ is a measure of how the target distribution differs from the latent space distribution. $\mathcal{L}_{Cluster}$ embeds a hierarchy of distances in the latent space, such that the distance between an anchor point and a valid point is minimized and vice-versa. As a result, the desirable points are all forced to form a cluster in one region and thus one can obtain an efficient sampling of the parameter space as depicted in the schematic diagram in Fig.~\ref{schematic2}.  

\paragraph{Performance}
The performance of SAFESPAM has been tested with both cMSSM and pMSSM scenarios with the 125 GeV Higgs Mass and Dark Matter Relic Density as the Observables \cite{Baretz:2023mra}.  For sampling from the latent space, Kernel Density Estimation (KDE) sampling has been used. Fig.~\ref{SAFESPAM_results} shows the performance in generating valid parameter points.


\paragraph{CHUNC(2)}
To avoid the potential information loss due to dimensional reduction in the mapping $T\rightarrow L$, two more methods have been considered, wherein the dimension of the latent space is the same as that of the GUT-scale parameter space (CHUNC) and the dimension of the latent space is one higher than that the GUT-scale space (CHUNC2). An additional constraint regarding the lightest supersymmetric particle (LSP) was added to test the performance of CHUNC and CHUNC2. Their performance can be seen in the RHS of fig.~\ref{SAFESPAM_results} \cite{Baretz:2023mra}. The reconstruction of the parameter space is more faithful in this case.

\subsubsection{Using AL to estimate higher dimensions}\label{sec:al_cmssm}
Unlike in the previous case (sec. \ref{sec:activelearn}), where AL was used as a classifier problem to determine the position of a point on one side of the decision boundary or the other, one can also use AL with a regressor to simulate the likelihood function. Without calling HEP packages to calculate the observable value, such approximate calculations can be very fast computationally. In sec. \ref{sec:activelearn}, random points were sampled and the interesting points---the ones which had the maximum uncertainty---were chosen. In the present example (ref. \cite{Ren:2017ymm}), the new points are chosen by selecting the points with a higher likelihood. However, in this case, a trained classifier is also used to recognize whether a point is physical or not. 

\begin{figure}[h!]
    \begin{subfigure}{\textwidth}
    \centering
        \includegraphics[width=0.95\textwidth]{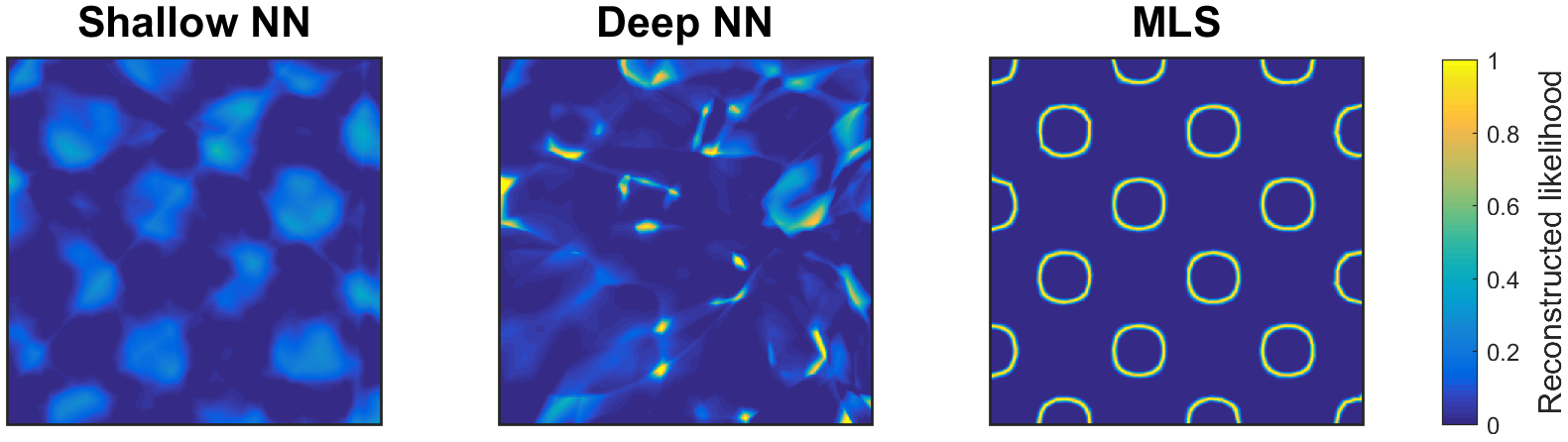}
    \caption{\small Reconstructed likelihood values for the eggbox toy model.}\label{eggbox}
    \end{subfigure}\\
    \begin{subfigure}{\textwidth}
    \centering
        \includegraphics[width=0.85\textwidth]{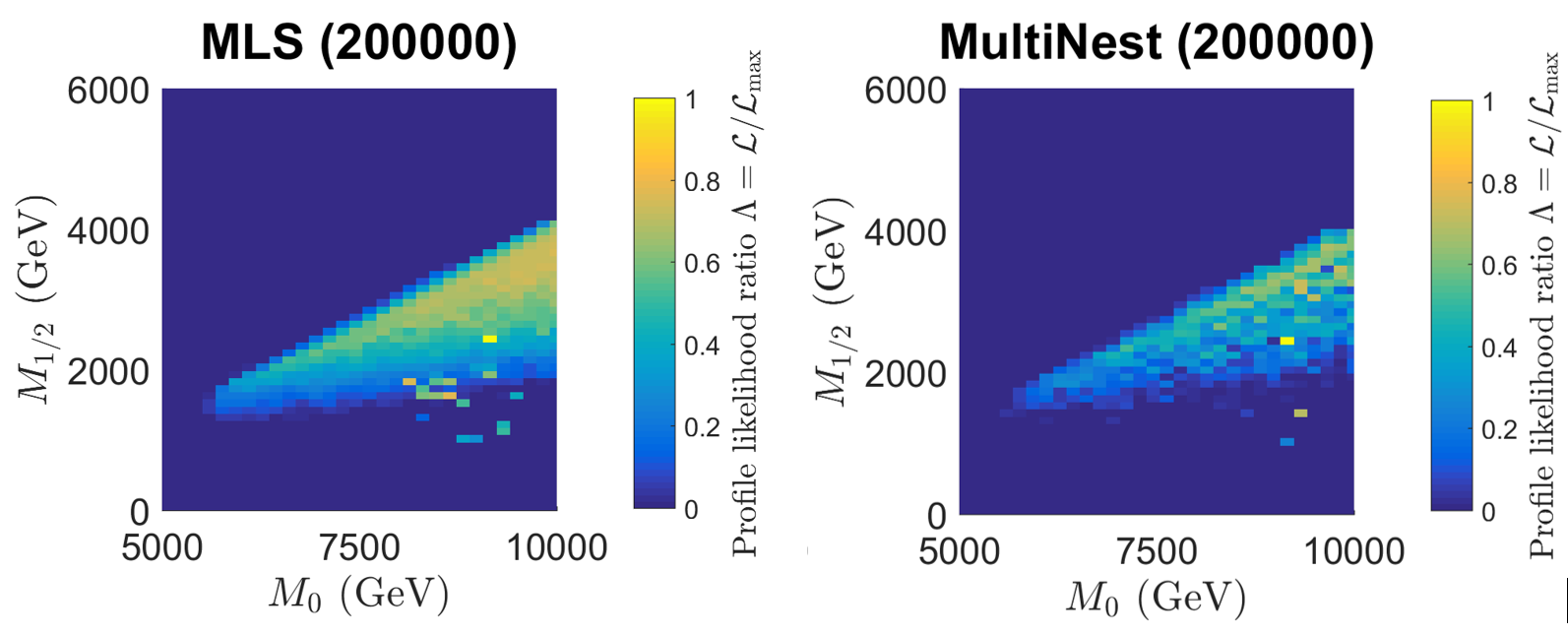}
    \caption{\small Prof. likelihood ratios for 20,000 points in $M_0 - M_{1/2}$ plane for MLS and Multinest.}
    \label{cMSSM}
    \end{subfigure}
    \caption{\small Reconstructed likelihood values and profile likelihood ratios for two different models discussed in sec. \ref{sec:al_cmssm}. Figures are taken from ref.~\cite{Ren:2017ymm}.}\label{fig:al_cmssm} 
\end{figure}

The framework, namely, the Machine Learning Scan (MLS; 3 to 4 hidden layers with 50-100 neurons per layer followed by ReLU) is applied to four different scenarios: a toy eggbox model, 2-D and 7-D variants of a quadratic model, cMSSM, and $R$-parity conserving SUSY. For the eggbox model, results are compared with a shallow NN (a single layer of 2000 neurons) and a deep NN (4 hidden layers with 100 neurons each), in predicting the likelihood values. Fig.~\ref{eggbox} clearly shows how the performance of MLS is superior \cite{Ren:2017ymm}. 

It can be seen that even with a lower number of sampling points, MLS outperforms traditional posterior sampling methods, even the faster NS ones (MultiNest), significantly. We show two representative cMSSM plots (one for MLS, the other for MultiNest) in fig.~\ref{cMSSM}. These depict the profile likelihood ratios in the $M_0$-$M_{1/2}$ plane \cite{Ren:2017ymm}.

\subsection{Sampling with simultaneous classification and regression}
\label{sec:mlassist}
\begin{figure}[h!]
  \centering
  \includegraphics[width=\textwidth]{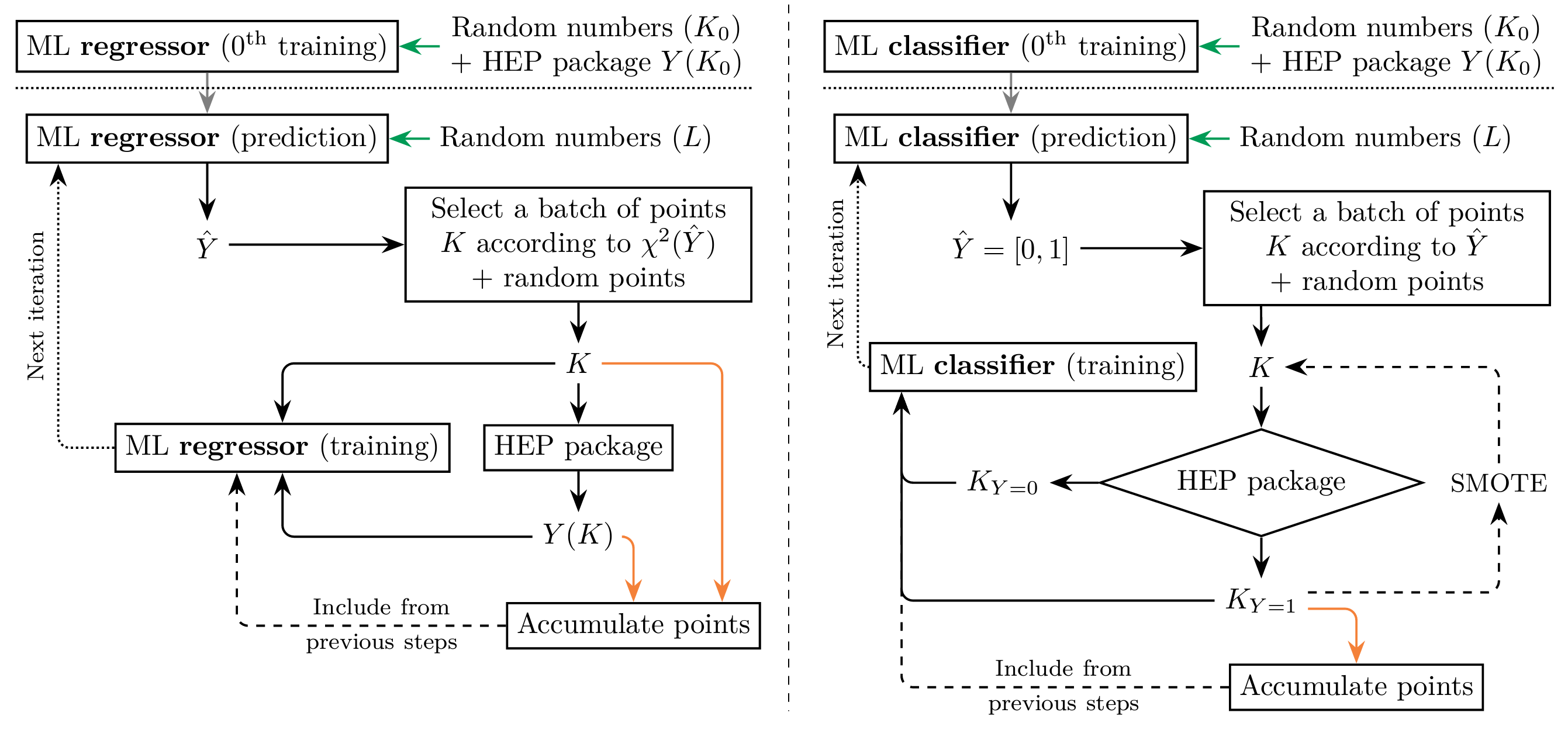}
  \caption{Illustrations depicting the iterative ML process for the regressor (left) and the classifier (right). The figure is taken from ref.~\cite{Hammad:2022wpq}.}
  \label{fig:2207_1}
\end{figure}

Likelihoods in BSM scenarios are famously computationally expensive---especially when they come with a large number of input parameters---often leading to poorly sampled regions. To circumvent this, ML-assisted algorithms with regression and classification have been proposed, e.g. ref. \cite{Hammad:2022wpq}. A training dataset of parameter values and corresponding High-Energy Physics calculations is used to train an NN and then the same NN is used to predict the parameter values that fit the data with certain precision. The network is used iteratively, improving the learning of the parameter space in every step. Apart from identifying a `good' or a `bad' point, the framework can also generate regions in the parameter space densely populated by points around the best-fit regions for a given set of observables.  

The ML model consists of a Deep Neural Network Classifier (DNNC) and a Deep Neural Network Regressor (DNNR). A linear activation function guides the regressor, while a sigmoid activation function aids the classifier. Loss functions are: binary cross-entropy-loss for the classifier and mean squared error for the regressor. To improve model performance, a technique called the synthetic minority oversampling technique (SMOTE) \cite{Chawla:2002dkc} has been employed. It generates synthetic samples for minority classes, ensuring a more balanced dataset. 

The following sequence summarises the approach \cite{Hammad:2022wpq}.
\begin{enumerate}
   \item Machine learning models (DNNC and DNNR) predict a broad set of random parameter values.
   \item A subset of points (K) is selected based on specific criteria, possibly including $\chi^2$ values or likelihood considerations.
   \item Selected points (K) undergo High-Energy Physics calculations, yielding actual results ($Y(K)$).
   \item Parameter values (K) and corresponding results ($Y(K)$) refine the machine learning models' training.
   \item The iterative process continues, accumulating points and integrating newly acquired data.
\end{enumerate}

The two flowcharts, depicted in fig.~\ref{fig:2207_1} \cite{Hammad:2022wpq}, illustrate the iterative process for the ML regressor (left) and the ML classifier (right) used for this purpose. Both begin with an initial 0th training step and advance to the first prediction step within the iterative process. Green arrows indicate where random parameter sets are inserted, while orange arrows indicate steps where points are accumulated in the sampled parameter space pool.

The effectiveness of this machine learning model is showcased with the example of a type-II higgs doublet model (2HDM). A thorough numerical study has been performed over seven free parameters of the 2HDM potential, namely, $\lambda_1$, $\lambda_2$, $\lambda_3$, $\lambda_4$, $\lambda_5$, $\tan \beta$, and the soft $Z_2$-breaking mass parameter $m_{12}$. The parameter space has been sampled within specific ranges ensuring vacuum stability. A wide range of observables has been used, which include the 125 GeV higgs boson mass, its signal strengths, production rates, and decay branching ratios in addition to electroweak precision observables (e.g., $S, T, U$ parameters) and $B$-meson decay measurements. Relevant theoretical and experimental constraints are also taken into account. The initial parameter ranges before training are strategically selected to encompass a broad spectrum of possibilities, adhering to necessary constraints.

 
A comparison among various sampling methods, namely, the ML classifier (DNNC), the regressor (DNNR), MCMC, and MultiNest further establishes the superiority of the ML algorithms. The efficiency analysis illustrates how each method converges to the desired number of points after sampling. This convergence also depends on certain factors, such as the initial number of target points and the use of SMOTE for data balancing. Fig.~\ref{fig:label} presents a visual representation of the efficiency of these four different methods \cite{Hammad:2022wpq}:  
\begin{figure}[]
  \centering
  \includegraphics[width=0.8\textwidth]{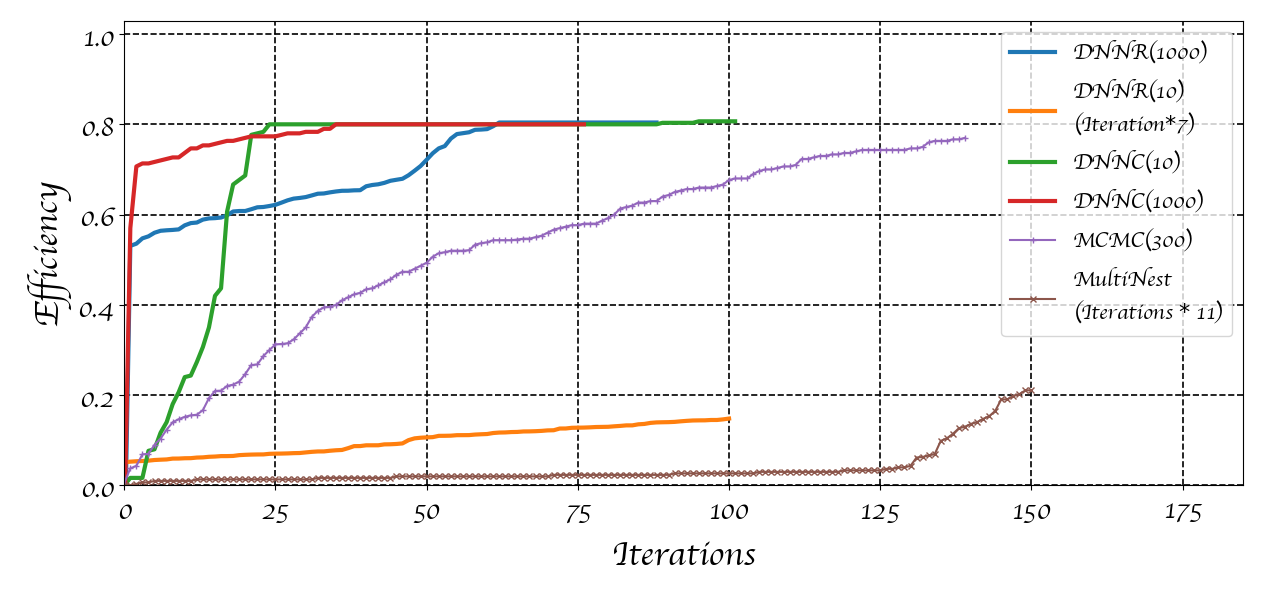}
  \caption{Efficiency across iterations is compared among various methods. Initial in-target points for each method are indicated in parentheses, except MultiNest. The figure is taken from ref.~\cite{Hammad:2022wpq}.}
  \label{fig:label}
\end{figure}

From the comparative analysis, it is observed that the ML models (DNNC and DNNR) show faster convergence than traditional techniques (MCMC and MultiNest), and the speed of convergence is highly dependent on the initial number of target points. In other words, this section explicitly shows why one may choose ML models over others in order to achieve fast exploration of parameter space in particle physics when there exists a significant complexity arising due to existing constraints and (or) high-dimensional parameter spaces.

\section{ML-assisted Nested Sampling}
\label{sec:MLAssist}
\subsection{Test Example}\label{sec:test_example}
\begin{figure}
  \centering
\begin{subfigure}{0.45\textwidth}
    \includegraphics[width=\textwidth]{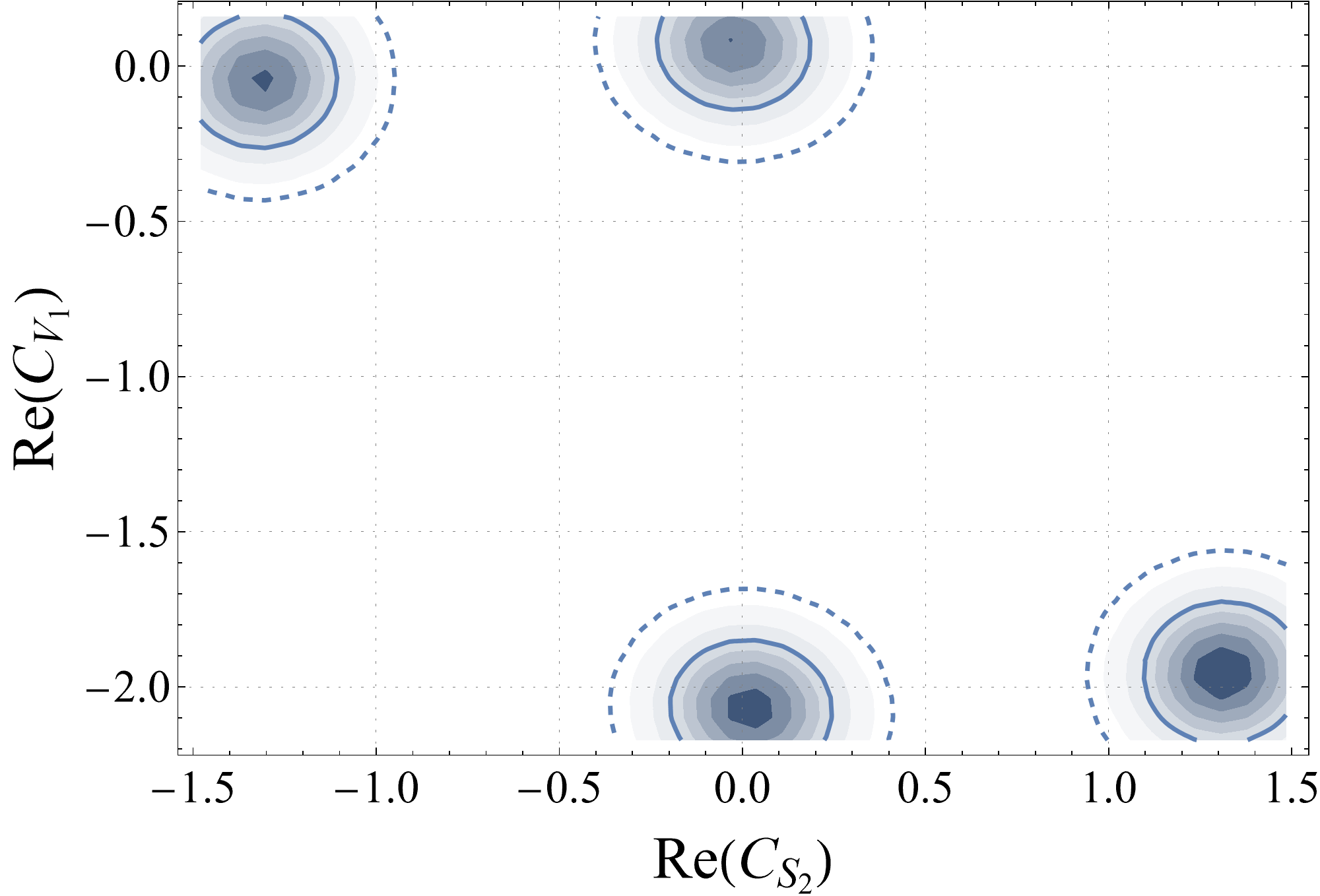}
    \caption{MCMC (M-H) Param-space.}
    \label{fig:comp1}
\end{subfigure}~~~~
\begin{subfigure}{0.45\textwidth}
    \includegraphics[width=\textwidth]{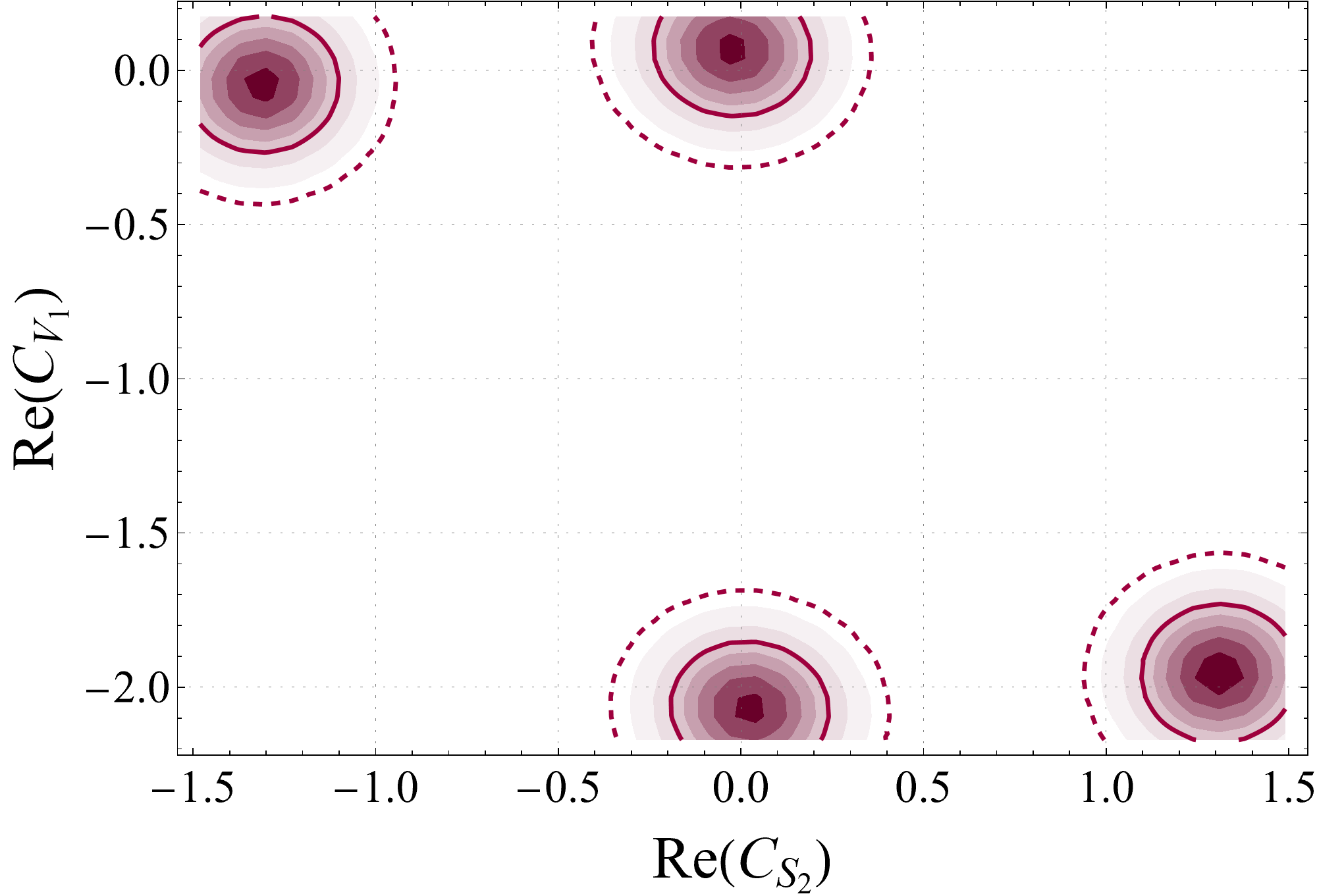}
    \caption{NS Param-space.}
    \label{fig:comp2}
\end{subfigure}\\
\begin{subfigure}{0.4\textwidth}
    \includegraphics[width=\textwidth]{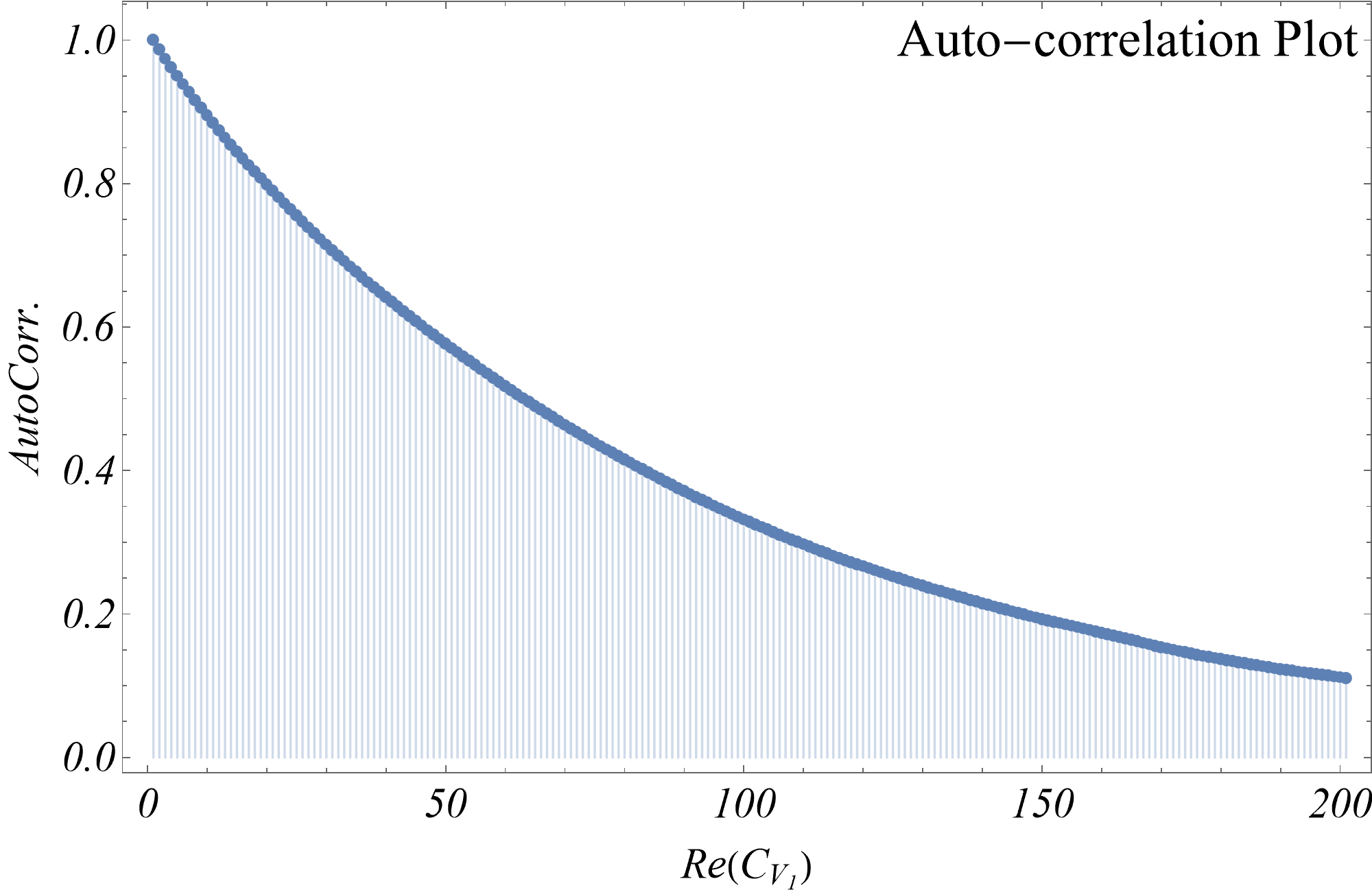}
    \caption{Auto-Correlation Plot.}
    \label{fig:comp3}
\end{subfigure}~~~~~~~~~
\begin{subfigure}{0.43\textwidth}
    \includegraphics[width=\textwidth]{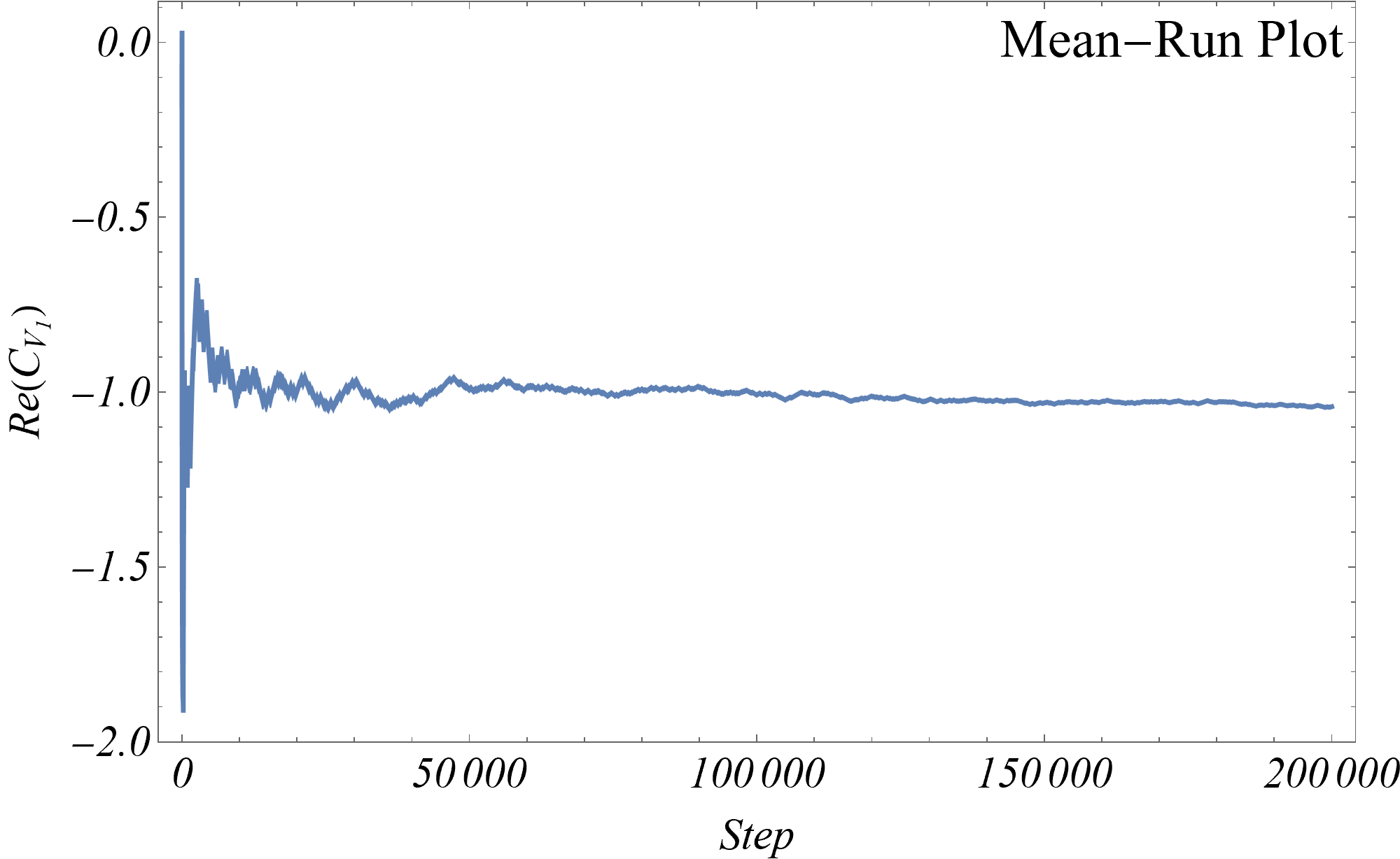}
    \caption{Running Mean Plot.}
    \label{fig:comp4}
\end{subfigure}
\caption{Comparison of traditional MCMC and NS algorithms for the example in sec. }
\label{fig:mcmcplotcomp}
\end{figure}

To demonstrate the upcoming ML-assisted methods in this section and compare the results at different points, we choose a real statistical data-combination problem with some effective NP operators present. Data is taken from the 2019 world average of $R(D)$ and $R(D^*)$ by HFLAV \cite{HFLAV:2019otj,HFLAVSemileptonic2019}. Details of the corresponding theory, including nuisance parameters (form factor parameters, among others) can be found in ref. \cite{Bhattacharya:2018kig}. Let us mention this much about the inference problem:
\begin{enumerate}
    \item There are 9 partially correlated data points.
    \item There are 2 completely unknown parameters, namely $\mathcal{R}e(C_{S_2})$ and $\mathcal{R}e(C_{V_1})$. The uniform priors used for these parameters are $\{-3, 3\}$.
    \item There are 5 nuisance parameters: $\rho_D^2$, $\rho_{D^*}^2$, $R_1(1)$, $R_2(1)$, and $R_0(1)$. The multi-normal prior of these parameters can be found in Table 6 of ref. \cite{Bhattacharya:2018kig}.
\end{enumerate}

\paragraph{Frequentist Results:} Let us mention some conventional frequentist results of the inference problem here, as a starting point, and a yardstick. Using the multi-normal distribution of the dataset, we construct a $\chi^2$-cost function and minimize it using a global optimization technique called Differential Evolution \cite{onwubolu2009differential}. With the assumption that the cost function follows a $\chi^2$ distribution of degrees of freedom (DOF) $= 9 - 2 = 7$, the reduced chi-square, $\chi^2_{Min}/DOF = 8.7709/7 = 1.253$ corresponds to a $p$-value of $26.95\%$. It turns out to be a multi-modal solution, with 4 distinct modes of the same height\footnote{The multi-modality is in the NP-dimensions. The nuisance planes are dominated by the prior and are uni-modal}, separated by much more than $3 \sigma$ confidence intervals.

\paragraph{MCMC Results:} To set a benchmark, we use a simple Metropolis-Hastings \cite{Hastings:1970} implementation of the MCMC algorithm in \textit{Mathematica} to obtain the parameter space. At first, a naive implementation of the algorithm, with the spread of the proposal distribution determined by the Hessian calculated at the single $\chi^2$-minimum, gets stuck around the said mode and is unable to find the other three. Then, using a thinning of 500 points and a modified spread of the proposal distribution, we save 200000 samples to obtain the full parameter space (fig: \ref{fig:comp1}). The obtained samples are still not perfectly \emph{iid}, as seen from the representative auto-correlation plot in fig: \ref{fig:comp3} but have more or less conveniently converged, as shown by the representative running mean plot in fig: \ref{fig:comp4}). High-density parts of the contours in fig: \ref{fig:comp1} also show that not all modes are equally sampled (they should be of the same probability). This run with a compiled likelihood function took $\sim 100$ mins for the run to complete. 

\paragraph{NS Results:} Next, the same parameter space is obtained by running a nested sampling algorithm with a simple cuboid region sampler, with 400 live points and a tolerance of $10^{-4}$ (this is a stopping criterion; see sec: 2.5 of ref. \cite{Ashton:2022}). With a weighted (importance) sample of $\sim 10000$ points, the run takes $\sim 7$ mins to finish. Fig: \ref{fig:comp2} shows that this still faithfully samples the parameter space. The logarithm of the evidence, $\log{Z} = 2.16 \pm 0.16$. We can increase the precision of both the evidence integral and the posterior by remembering that both scale as $\mathcal{O}(1/\sqrt{n_{live}})$, while the run time scales as $\mathcal{O}(n_{live})$.

\paragraph{Adding (pseudo)-constraints and Slow-likelihood:} To make our test problem simulate the constrained parameter spaces typical to the problems in this domain, we choose a (pseudo)-constraint $Re(C_{S_2}) \leq -0.6 \lor 0.6 \leq Re(C_{S_2})$, to ensure a large disconnect between two parts of the remaining allowed region. Another typical characteristic of the BSM parameter-search problems is the bottleneck of large, complicated, and hence, computationally expensive likelihoods. A large number of individual likelihood estimations are needed for a reliable and sufficiently converged result, rendering the parameter space intractable. This is one of the main motivations for introducing the present technique. To simulate this effect for our test problem, we randomly add some seconds (between 30 sec and 1 min) to every likelihood estimation. We will call this the `Slow-likelihood'.

\paragraph{NS on newly defined likelihoods:} We run a standard NS with cuboid region sampler ($n_{Live} = 700$, weighted sample-size $= 17800$, final tolerance $= 10^{-4}$, run-time is 3 hr $9\frac{1}{2}$ min) to get $log(Z) = 0.19 \pm 0.14$. Repeating the runs with different samplers (fast-ellipsoid region, oblique cuboid region, and random-walk sampler) does not change $log(Z)$ by any considerable amount. We then run the `slow' version of the constrained likelihood to get an estimate of the increase in run-time. To get even a tolerance of $\sim 10$, it took $> 10$ hrs. This gives us a timeline to compare, and hopefully, beat.

\begin{figure}[]
  \centering
  \includegraphics[width=0.6\textwidth]{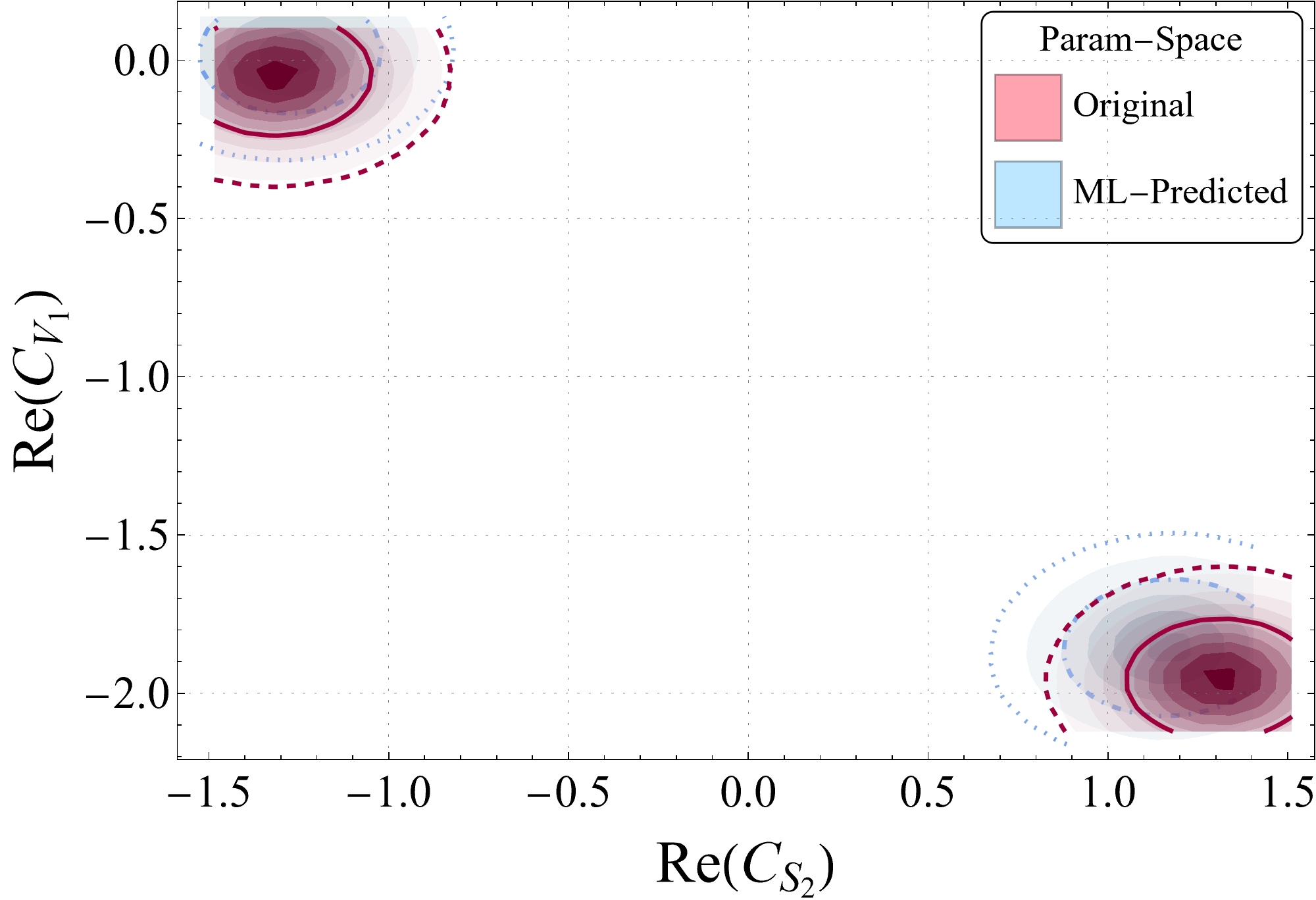}
  \caption{Comparison of a pure ML prediction with the actual likelihood.}
  \label{fig:mlNSCompPlot2D1}
\end{figure}

\subsubsection{Method}

The typical BSM parameter estimation problem has two parts, finding a way to apply external constraints, either before or after likelihood estimation, and calculating the posterior. Constraints, applicable only after the likelihood calculation, are more time-consuming. To alleviate all these problems in one go, we propose \emph{`ML-assisted NS'}. In essence, it is close to an active learning problem that simultaneously finds the domain boundary classifier and a trained regressor. In contrast to other AL methods shown in this paper, it performs a pure NS run during the training process, providing us with a well-defined posterior and a handle on the evidence in favor of the model in question. The somewhat generic name is deliberately chosen to incorporate later developments to all internal algorithms. For now, it works in the following manner:

\paragraph{Create dataset:} A reasonably large set of random points are selected from the $d_p$ dimensional prior space (for the test case, $d_p = 2+5 = 7$), each represented by a $d_p$-dimensional vector. For each of them, a $d_o$-dimensional vector, with the theoretical prediction of observables as its elements, is then created (test case: $d_o = 9$). We then check the constraints on each of them (depending on the problem, this check may need to be run before calculating the observables). A single dataset (key-value association) is then crafted with the $d_p$-dimensional vectors as keys and two-part values: the first part is either $0$ or $1$, denoting whether the point in question violates the constraints or not; the second part, which is a non-empty list for only points with $1$ in the first part, is the corresponding $d_o$-dimensional vector. 

\begin{figure}[]
  \centering
  \includegraphics[width=0.9\textwidth]{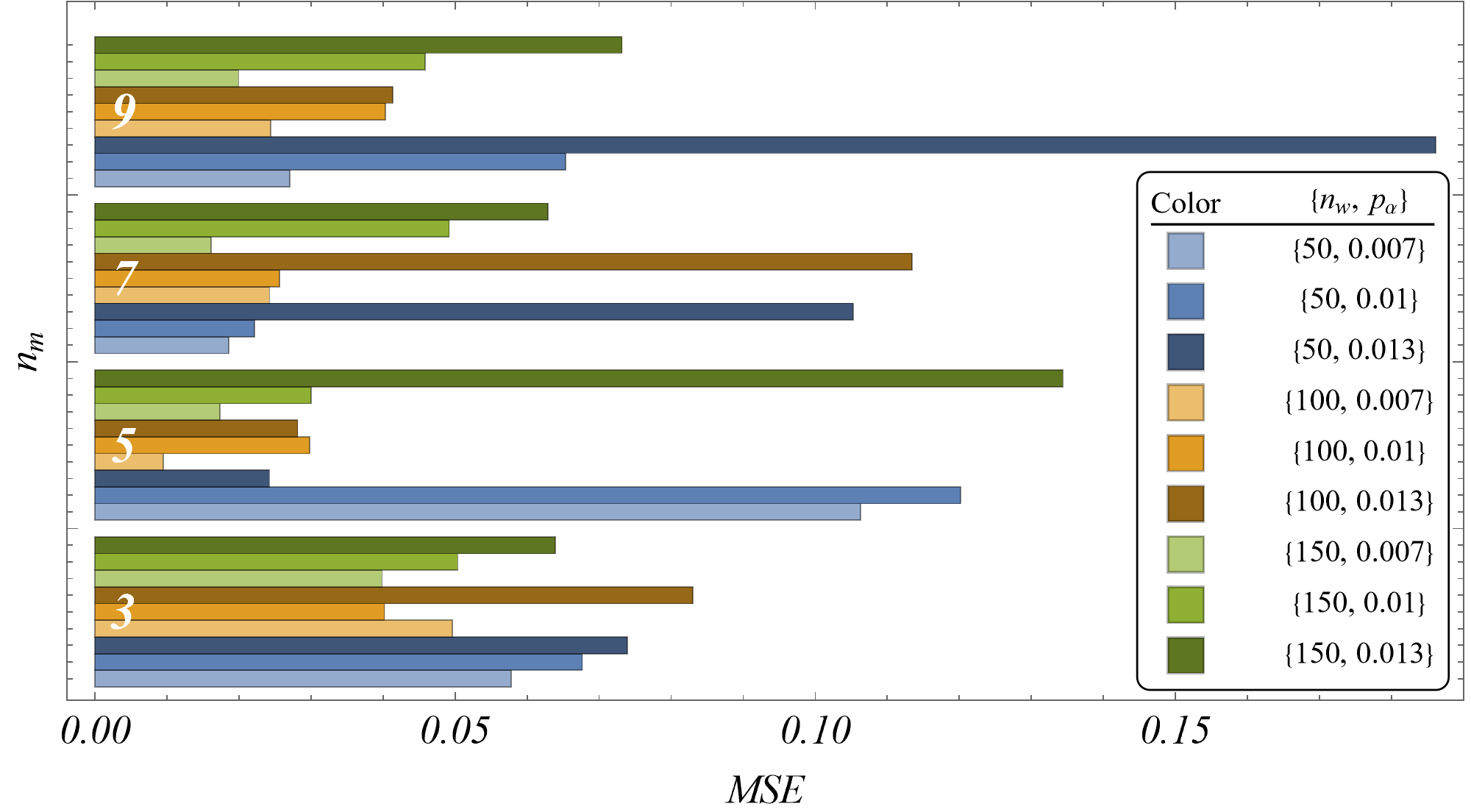}
  \caption{$n_m$, $n_w$, and $p_{\alpha}$ are, respectively, the number of modules, the width of each module, and the dropout probability of the SNN in question.}
  \label{fig:measurementChart1}
\end{figure}

\paragraph{Pre-training:} We need two ML algorithms for our method: one classifier to predict the decision boundary (a binary classifier mostly, but depending on the problem, it could test for multiple conditions, translating to a multi-class classifier) and one regressor to predict the observable-vector for points allowed by the classifier. In principle, any ML algorithm can work for these---our code has provision for that but, for our purposes, these are two variable-structure `Self-normalizing Neural Networks (SNNs)', each of which has $n_m$ modules, followed by a linear layer of required length (the classifier would have a soft-max layer after it). Each module of an SNN has two layers - one layer of $n_w$ artificial neurons with RELU-activation function and a specialized $\alpha$-dropout layer of dropout probability $p_{\alpha}$. These three -- $p_{\alpha}, n_w$, and $n_m$ can be tuned as hyper-parameters. These SNNs are then trained with the created dataset. This gives us a starting point for the NS algorithm.

Before we get into posterior determination, let us talk about the training in a little more detail. We could very well use the likelihood or the negative of it (modified to represent $\chi^2$) for the target of the training dataset, which we tried before. Due to loss of information (dimension of the output shrinks from $d_o$ to 1), that training has a lot of uncertainty and was deemed unsuitable for our purposes. In this sense, the `trained' combination of the classifier and regressor represents the NP model within error bars for the observables in question. Using these predictions to calculate the predicted likelihood $\mathcal{L}_{pred}$ makes it more stable and simulates the likelihood better.

\begin{figure}[]
  \centering
  \includegraphics[width=0.75\textwidth]{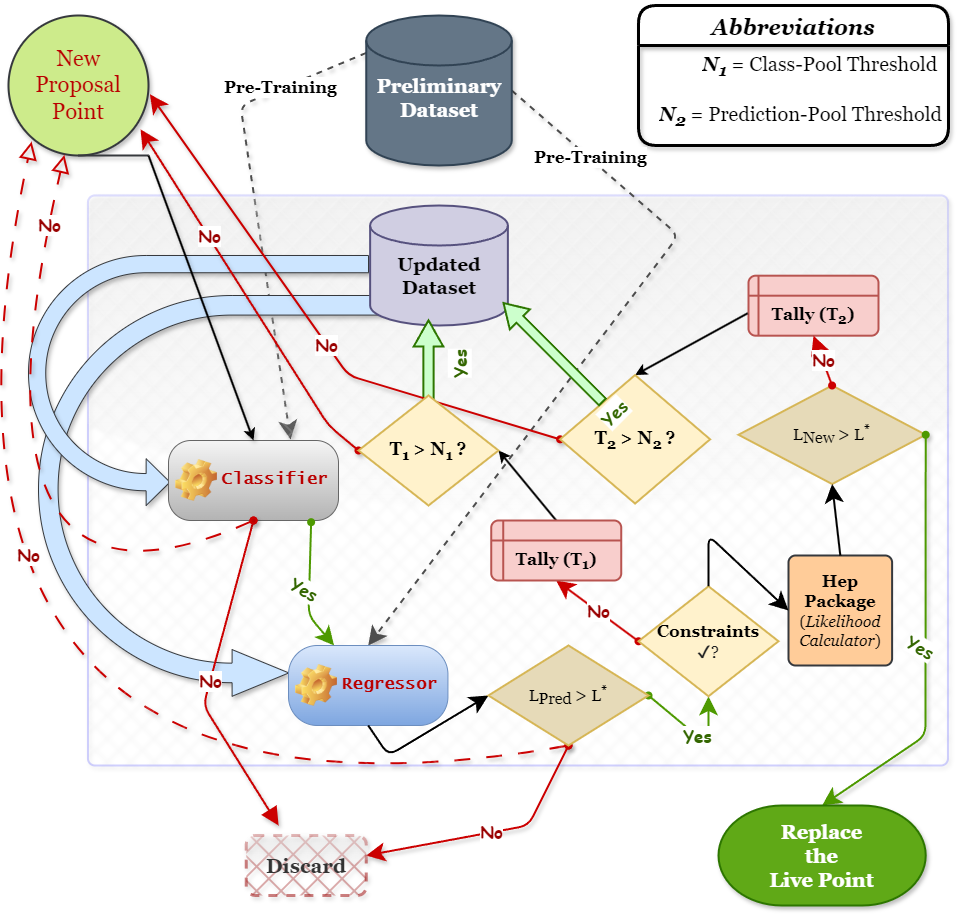}
  \caption{ML-assist in one iteration of the NS run.}
  \label{fig:ML-NS_Internal}
\end{figure}

We could also do away with any sampling method altogether and spend more time generating training data and then training a `better' set of classifiers and predictors. Such an attempt was made while varying the hyper-parameters, using an `infinite pool' of data (i.e., a generating function for the observables), until the Coefficient of Determination ($R^2$) reaches stability for the regressors. We test various performance metrics, e.g., $R^2$, standard deviation as well as mean-square of the residuals (MSE) to find the best configuration of the SNN to train for this particular problem. Fig. \ref{fig:measurementChart1} shows the bar chart of MSE for all these configurations. We see that the MSE is lowest for the setup: $(n_m, n_w, p_{\alpha}) = (5, 100, 0.007)$, i.e., a wide but short network with less dropout regularization works best. It also demonstrates that while a certain level of complexity is needed for unbiased prediction of the problem, bigger is not always better for neural network configurations (a direct result of the principle of parsimony). The trained SNN of this configuration then is used as a replacement for the NP model here and is used to find the `predicted' posterior by using a standard NS algorithm (with the same run-configuration, i.e., $n_{live}$, tolerance, etc.). It took $> 20$ hrs to run up to a tolerance of $0.5$ with $log(Z) = 0.94 \pm 0.13$. In addition to the obvious differences found after comparison with the results at the end of sec. \ref{sec:test_example}, the main problem is that the obtained parameter space is biased, albeit by a small amount. Prediction for this bias for other problems is not possible. This is why we refrain from completely replacing the actual model with an ML one. 

\paragraph{ML-assisted NS:} The solution, for us, is to use ML to widen the runtime bottleneck as much as possible. It is clear from eqs \ref{eq:Z1} -- \ref{eq:Z3} and algorithm (Fig.~\ref{algo:NS}) that the most important step of any NS algorithm is to replace the point with the lowest likelihood $L^*$, among the pool of $n_{live}$ points, with another one with $L > L^*$, randomly sampled from the prior. This sampling can be done in many ways, e.g., random walk (MCMC), randomly from a region, or splice sampling. Whichever is the case, this search requires tens, if not hundreds, of likelihood evaluation. This is that bottleneck we talked about. We propose to use an ML-assisted sampler for this purpose, which, will use (after a regular number of steps) the actively trained regressor to find candidates for the replacement point. It works in tandem with the classifier at each step, to ensure that all points pass the constraints of the NP model. 

Fig. \ref{fig:ML-NS_Internal} demonstrates the main parts of the algorithm. A preset number of points are proposed randomly, which pass through the classifier (trained with the preliminary dataset). Those that return 0 or `false' in this binary classifier, are discarded (\textit{first speed-up}). If no points satisfy the constraints, we restart the loop (shown in the figure with dashed arrows labeled `No'). Constraint-satisfied points are then fed to the regressor to obtain an ML-predicted likelihood estimate for them. Points that do not satisfy $L_{pred}>L^*$ are also discarded (\textit{second speed-up}). If no points pass the check, we restart the loop. Constraints are then actually tested on any remaining points. Those failing the test are saved and tallied ($T_1$), and the rest progress to be processed by some external HEP package to generate the actual likelihood (in this test case, it is the slow-likelihood). The points failing the check $L_{new}>L^*$ are saved and tallied separately ($T_2$); one point is randomly chosen from those passing the check and is used in replacing the live point corresponding to $L^*$. 

As can be checked from the schematic (fig. \ref{fig:ML-NS_Internal}), the tallies $T_1$ and $T_2$ are for those points, for which the SNNs made false positive predictions. Using two different large numbers ($N_1$ and $N_2$) as thresholds, we interrupt the NS algorithm to append the training dataset with the tallied and saved points (with their actual likelihoods and constraint relation) and re-train the classifier and the regressor. This is the AL part of the algorithm.

We know that NS slows down following a high power law with time, as it becomes increasingly harder to get a higher density point near the modes. This becomes even more painful with time-consuming likelihoods. With the ML-assist, this slowdown happens linearly. Along with the speed-ups mentioned earlier in this section, ML-assisted NS gives a considerable boost to the speed of the numerical calculation. Still, the resulting $log(Z)$ and parameter spaces are identical to the normal NS result---as they were obtained by calculating the actual likelihoods.

\subsection{Comments}
Further tests with diverse cases are needed to comment on the actual consistent speed-up factor of this method, but for the test case here, it is a factor of at least $\sim 8$ and at most $\sim 12$, w.r.t. the actual slow-likelihood times. Varying the calculation time of the slow-likelihoods slows the normal NS down by the same factor. The slowdown of the ML-assisted case is much smaller, as most of the iterative $L$-s are SNN-predicted $L_{pred}$-s, taking essentially no time to calculate.

Though the methods discussed in subsection \ref{sec:mlassist} may seem quite similar to the one in sec. \ref{sec:MLAssist}, there are intrinsic differences that render their effect and utility largely different. ML-Assisted NS calculates the \textit{Evidence} of the posterior and samples it while training the classifier-regressor combo. The one in sec. \ref{sec:mlassist}, on the other hand, just trains the NNs. Points accumulated during the training are not used in drawing confidence intervals. Using SMOTE during the training of the classifier, as in sec. \ref{sec:mlassist}, is something the ML-Assisted NS could benefit from.

\section{Summary}
\label{sec:summary}
In this article, we have summarised the effort of the particle physics community in addressing a pertinent issue, that of sampling intractable parameter spaces of various BSM scenarios subjected to the available experimental data using machine learning tools. Traditionally, posterior inference is generated using statistical tools such as MCMC which are computationally expensive and are not very efficient in identifying all the solutions in a multi-modal parameter space. Nested sampling can be used to circumvent the multi-modality issue, but the algorithm gets slower as it approaches higher likelihood regions. Convergence of the algorithm thus remains time-consuming. The use of supervised learning has gained momentum to circumvent the difficulty of handling large parameter spaces. We have summarised some major efforts in this direction. ML has been used extensively in identifying `good' and `bad' points and in order to obtain the decision boundaries that separate those regions. Training through an infinite pool of data proves to be less effective compared to active learning which performs the training in an iterative way at each step. Algorithms based on active learning achieve better accuracy with a much smaller training size and one can sample the region around the decision boundary very efficiently. One can use generative autoencoders to reduce the dimension of a higher dimensional parameter space to a lower dimensional latent space that can effectively cluster the desirable points in one region. These tools, although efficient, cannot generate the posterior inference which paints a more complete picture of the parameter space in question. A bottleneck to this approach is the likelihood calculation. One can train a network to predict higher likelihood points instead of using HEP packages to compute the likelihood function at every point. Hardly any attempts have been made in this regard. To this end, we have proposed a framework where a trained neural network can predict higher likelihood points for a nested sampling algorithm. The neural network can be trained iteratively so that as the process reaches higher likelihood regions, the network gets more efficient and accurate in its predictions and in the process helps the code to converge on the high likelihood regions much faster. We have argued the effectiveness of our approach with a toy example.

  \vspace{+1cm}

 \noindent
\textbf{Acknowledgments:} RB, SM and SKP would like to acknowledge support from DST-SERB, India (grant order no. CRG/2022/003208). This research was supported in part by the International Centre for Theoretical Sciences (ICTS) through the workshop ``Statistical Methods and Machine Learning in High Energy Physics" (code: ICTS/ml4hep2023/08).

 \vspace{+1cm}

\noindent
\textbf{Data availability statement:} No data associated in the manuscript.


\bibliography{ml-bibliography}

\end{document}